\documentclass[11pt]{article}
\usepackage[english]{babel}
\usepackage[utf8x]{inputenc}
\usepackage[T1]{fontenc}

\usepackage{caption}

\usepackage{booktabs} 

 \usepackage{fullpage}

 \usepackage{tabularx}
 \usepackage{xcolor}
 \usepackage{multirow}
 \usepackage{mathtools}
 \usepackage{nicefrac}
 \usepackage{amsfonts,amssymb,amsthm}
 \usepackage{url}
 \usepackage{lipsum}
 \usepackage{commath}
 \usepackage{algorithm}
 \usepackage[inline]{enumitem}
 \usepackage[justification=centering]{caption}
 \usepackage[noend]{algpseudocode}
 \usepackage{setspace}

\usepackage{subcaption}
\usepackage[super]{nth}
\usepackage{color-edits}
\addauthor[Kartik]{kl}{blue}
 \algnewcommand\algorithmicparfor{\textbf{for}}
\algnewcommand\algorithmicpardo{\textbf{do in parallel}}
\algrenewtext{ParFor}[1]{\algorithmicparfor\ #1\ \algorithmicpardo}
\algrenewtext{EndParFor}{}
\algdef{SnE}[PARFOR]{ParFor}{EndParFor}[1]{\algorithmicparfor\ #1\ \algorithmicpardo}%

\begin{document}
\title{GPOP: A scalable cache- and memory-efficient
framework for Graph Processing Over Partitions}
\author{Kartik Lakhotia\quad Sourav Pati\quad Rajgopal Kannan\quad Viktor Prasanna\\
University of Southern California\\
\{klakhoti, spati, rajgopak, prasanna\}@usc.edu}
\date{}
\maketitle

\begin{abstract}
Past decade has seen the development of many shared-memory graph processing frameworks, 
intended to reduce the effort of developing high performance parallel applications. However many of these frameworks, based on Vertex-centric or Edge-centric paradigms suffer from several issues, such as poor cache utilization, irregular memory accesses, heavy use of synchronization primitives and theoretical inefficiency, that deteriorate overall performance and \looseness=-1scalability.

Recently, we proposed a cache and memory efficient partition-centric paradigm for computing PageRank~\cite{pcpm}. In this paper, we generalize this approach to develop a novel 
Graph Processing Over Partitions~(GPOP) framework that is cache-efficient, scalable and work-efficient. 
GPOP induces locality in memory accesses by increasing granularity of execution to vertex subsets called 'partitions', thereby dramatically improving the cache performance of a variety of graph algorithms.
It achieves high scalability by enabling completely lock and atomic free computation. GPOP's built-in analytical performance model enables it to use a hybrid of source and partition-centric communication modes in a way that ensures work-efficiency  each iteration, while simultaneously boosting high bandwidth sequential memory accesses. Finally, the GPOP framework is designed with programmability in mind. It completely abstracts away underlying parallelism and programming model details from the user, and provides an easy to program set of APIs with the ability to selectively continue the active vertex set across iterations. Such functionality is useful for many graph algorithms but not intrinsically supported by the current frameworks.
	
We extensively evaluate	the performance of GPOP for a variety of graph algorithms, using several large datasets. We observe that GPOP incurs up to $9\times$, $6.8\times$ and $5.5\times$  less L2 cache misses compared to Ligra, GraphMat and Galois, respectively. In terms of execution time, GPOP is upto $19\times$, $9.3\times$ and $3.6\times$ faster than Ligra, GraphMat and Galois respectively. 
\end{abstract}

%
%

\section{Introduction}\label{sec:intro}

Real world problems arising in web and social networks, transportation networks, biological systems etc. are often modeled as graph computation problems. Applications in these domains generate huge amounts of data that require efficient large-scale graph processing. To this purpose, many distributed frameworks have been proposed to process very large graphs on clusters~\cite{pregel,giraph,graphlab,comblas}. However, because of the high communication overheads of distributed systems, even single threaded implementations of graph algorithms have been shown to outperform many such frameworks running on several machines~\cite{cost}. 


The growth in DDR capacity also allows large graphs to fit in the main memory of a single server. Consequently, many frameworks have been developed for high performance graph analytics on multicore platforms~\cite{ligra, graphmat, galois, xstream,polymer,grazelle}. However, multi-threaded graph algorithms may incur race conditions and hence, require expensive synchronization~(atomics or locks) primitives that can significantly decrease performance and scalability. Furthermore, graph computations are characterized by large communication volume and irregular access patterns that make it challenging to efficiently utilize the resources even on a single machine~\cite{lumsdainechallenges}.

Recent advances in hardware technologies offer potentially new avenues to accelerate graph analytics, in particular, new memory technologies, such as Hybrid Memory Cube~(HMC) and scratchpad caches. However, many graph analytics frameworks are based on the conventional push-pull Vertex-centric processing paradigm~\cite{ligra, polymer,grazelle,topushortopull},  which allows every thread to access and update data of arbitrary vertices in the graph. Without significant pre-processing, this leads to unpredictable and fine-grained random memory accesses, thereby decreasing the utility of the wide memory buses and deterministic caching features offered by these new architectures. Some frameworks and application specific programs\cite{xstream,gridgraph,hpec,pcpm} have adopted optimized edge-centric programming models that improve access locality and reduce synchronization overhead. However, these programming models require touching all or a large fraction of edges of the graph in each iteration, and are not work optimal for algorithms with dynamic active vertex sets, such as bfs, seeded random walk etc. A work inefficient implementation can significantly underperform an efficient serial algorithm if the useful work done per iteration is very small.

 

In this paper, we discuss GPOP: A Graph Processing Over Partitions framework, that enhances cache-efficiency and memory performance by executing graph algorithms at a granularity of cacheable vertex subsets called \textit{partitions}.
The execution sequence in GPOP consists of steps of message exchange~(scatter) between partitions, and message processing~(gather) to compute new vertex values. GPOP also utilizes a hierarchical active list structure to ensure that work done in each iteration is proportional to the number of active edges, and a novel dual communication mode technique that enables completely sequential memory accesses while scattering messages from high workload partitions. Thus, GPOP extracts maximum performance from the memory hierarchy while guaranteeing theoretically efficient implementations. Note that GPOP uses a lightweight index-based partitioning and does not rely
on any sophisticated (and often time consuming) partitioning or vertex reordering schemes to enhance locality. 


 
GPOP provides exclusive ownership of a partition to one thread to avoid unnecessary synchronization and achieve good scalability. We further utilize this property of GPOP along with the shared-memory model of multicore systems, to enable interleaving of Scatter and Gather phases. This leads to asynchronous update propagation within each iteration and faster convergence for several algorithms.
The private ownership also allows development of 
an easy to program set of APIs, that completely abstract away the underlying parallelism from the user and
ensure correctness of the parallel implementation without requiring the user to lock or atomically update the vertex data. This is in stark contrast with many popular frameworks~\cite{ligra, polymer, grazelle} where the burden of ensuring correctness of a parallel program is left to the user. GPOP also facilitates flexible manipulation and {\it selective continuity} of the active vertex set across iterations, which is not possible with  existing frameworks.

The major contributions of our work are as follows:
\begin{enumerate}
	\item We propose the Graph Processing Over Partitions (GPOP) framework
    with novel optimizations that \begin{enumerate*}[label={(\alph*)}]
		\item improves cache performance and achieves high DRAM bandwidth, 
		\item minimizes synchronization and promotes fast convergence, and,
		\item ensures work efficiency of a given algorithm.
	\end{enumerate*}
	\item GPOP provides an easy to program set of APIs allowing selective continuity in frontiers across iterations. This functionality is essential for many algorithms such as Nibble, Heat Kernel PageRank~\cite{nibble} etc., but is not supported intrinsically by the current frameworks.

\item GPOP ensures parallel correctness without the use of synchronization primitives, and demonstrates good scalability.
	
	\item We evaluate the performance of GPOP by executing a variety of graph algorithms on large real-world and synthetic datasets. We show that GPOP incurs significantly less cache misses and execution time compared to baselines -- Ligra, GraphMat and Galois.
\end{enumerate}


%
%
%
\section{Graph Processing Challenges}\label{sec:challenges} 



Typically, graph applications execute iteratively, with a {\it frontier} of active vertices in each iteration, until a halting criteria is met. The performance of a graph analytics framework depends on the cache efficiency, DRAM communication, use of synchronization primitives and theoretical efficiency of its underlying programming model. 
We say that a parallel implementation is \textit{theoretically efficient} if the amount of work done in an iteration is proportional to the number of outgoing edges from active vertices i.e active edges (for single iteration execution and some applications, such as BFS, this is equivalent to being work-efficient). Note that theoretical efficiency implies the work done is optimal for the given algorithm, within constant factors. In this section, we illustrate several challenges faced by current frameworks, in regard with the aforementioned aspects.

Vertex-centric~(VC) programming is one of the most popular approaches used in graph processing~\cite{pregel}. Shared-memory frameworks~\cite{ligra, polymer, grazelle} typically execute an iteration using either of the following:~(ref. Algorithm~\ref{alg:pushpull})

$\bullet$
\textit{Push} VC -- where the set $V_a$ of active vertices (\textit{frontier}) is divided amongst worker threads who modify the attributes of their out-neighbors. In general, synchronization primitives such as atomics are required to ensure correctness of concurrent threaded access to common out-neighbors; this can limit the scalability of the program. While atomics can enable limited number of simple actions (compare and swap, fetch and add), applications performing a complex set of operations may require more expensive \textit{locks} on vertices. 

$\bullet$
\textit{Pull} VC -- where threads traverse the in-edges of vertices and modify their private attributes, requiring no synchronization. However, since edges from active and inactive neighbors are interleaved, all edges to the vertices in the 1-hop neighborhood of current frontier need to be processed. Since finding unexplored neighbors is the only goal in BFS, bottom-up BFS (pull direction) instead probes the in-edges of all vertices not yet visited.


\begin{algorithm}[htbp]
    \centering
	\caption{Vertex-centric push and pull methods}
	\label{alg:pushpull}
	\begin{algorithmic}[1]
\Statex{$N_i(v)$/$N_o(v)$: in/out neighbors of vertex $v$} 
\Statex{$val[v]$: attribute of $v$ (eg. distance in SSSP, parent in BFS)}
		\Function{push}{$G(V,E), V_a$}
			\ParFor {$v \in V_a$}
					\ForAll {$u\in N_o(v)$}
						\State{$val[u] = \text{updateAtomic}(val[v], val[u])$}
					\EndFor
			\EndParFor
		\EndFunction
		\Function{pull}{$G(V,E)$}
			\ParFor {$v \in V$}
				\ForAll {$u\in N_i(v)$}
					\State{$val[v] = \text{update}(val[v], val[u])$}
				\EndFor
			\EndParFor
		\EndFunction		
	\end{algorithmic}
\end{algorithm} 

\begin{figure}[htbp]
    \centering
	\includegraphics[width=0.7\linewidth]{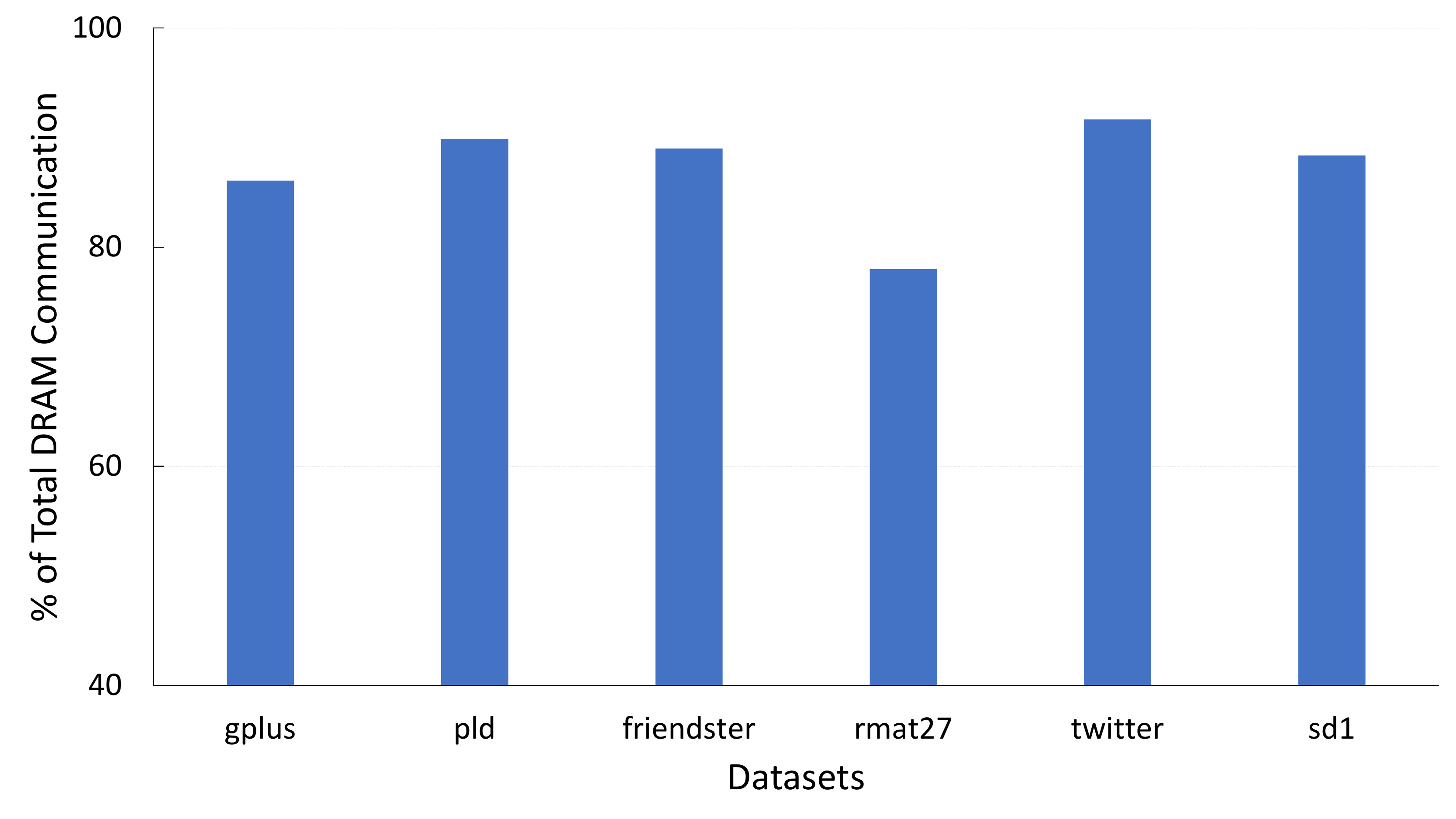}
	\caption{DRAM Traffic generated by random accesses to vertex data in PageRank}
	\label{fig:vertexTraffic}
\end{figure}

Most frameworks store the graph in a Compressed Sparse format (CSR or CSC) \cite{lumsdaineboost} which allows efficient sequential access to the edges of a given vertex. However, acquiring values of neighboring vertices requires fine-grained random accesses as neighbors are scattered. For large graphs, such accesses increase cache misses and DRAM traffic, becoming the bottleneck in graph processing. Figure~\ref{fig:vertexTraffic} is an illustrative example. Here, $>75\%$ of the total DRAM traffic in a PageRank iteration (in a hand-optimized code) is generated by the random vertex value accesses.

To improve locality of memory accesses, Edge-Centric~(EC) programming and blocking techniques \cite{xstream, spmv, gridgraph} are also used. In EC, vertex values are written on edges in a semi-sorted manner (analogous to bucket sort) to increase spatial locality and cache-line utilization. Blocking restricts the range of vertices accessed by each thread and thus, improves cache hit ratio. While such methods improve cache performance, they still incur coarse-grained random accesses limiting the bandwidth of CPU-DRAM communication. Recently, we proposed a Partition-Centric Programming Methodology (PCPM) \cite{pcpm} to avoid random DRAM accesses and sustain high bandwidth for PageRank computation. However, all of these approaches~\cite{spmv, gridgraph, xstream, pcpm} iterate over all vertices and edges in the graph, making them theoretically inefficient and extremely expensive for executing iterations with very few active vertices. The improved locality in these methods further comes at the cost of fully synchronous vertex value propagation that reduces the convergence rate for certain graph algorithms, such as single source shortest path computation.
 \section{Graph Processing Over Partitions}\label{sec:prog}

\begin{figure}[htbp]
	\centering
	\includegraphics[width=0.6\textwidth]{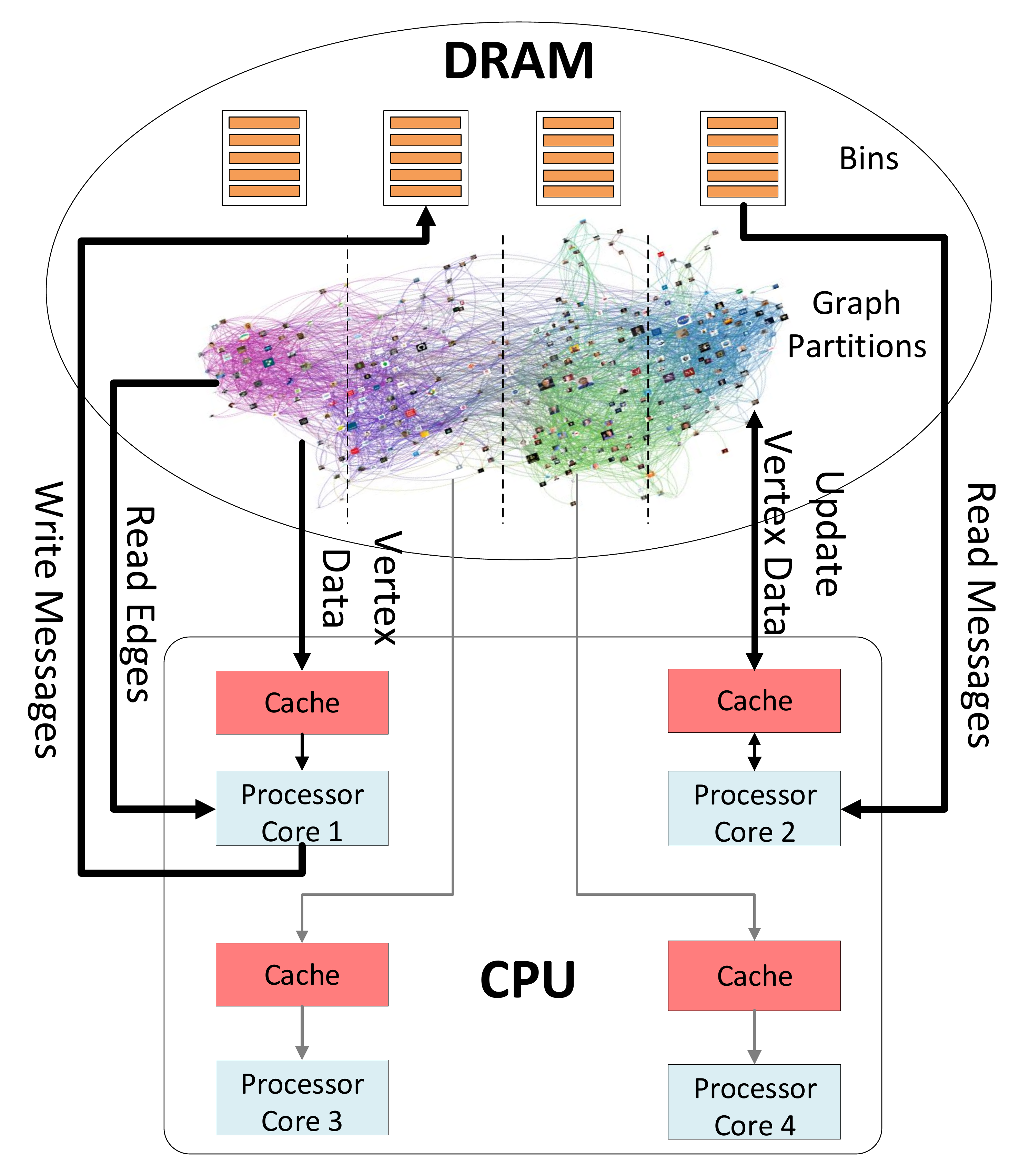}
	\caption{Conceptual demonstration of two phases in GPOP - Core $1$ is writing messages in Scatter phase and Core $2$ is processing them in Gather phase. Note that in the actual implementation, all threads are in the same phase at any given instant of time.}
	\label{fig:pcpm_model}\vspace{-2mm}
\end{figure}

In this paper, we describe GPOP: a Graph processing framework that enhances cache and memory performance, while {\it guaranteeing} theoretically efficient execution of a variety of graph algorithms. GPOP generalizes the partition-centric paradigm developed in \cite{pcpm} which was restricted only to PageRank. Dividing the vertex set into {\it cacheable} disjoint parts, GPOP implements each iteration of an algorithm in 2 phases:
\begin{enumerate}[leftmargin=*]
\item	Inter-partition communication~(\textit{Scatter}) $\rightarrow$ In this phase, the out edges of active vertices in a partition are streamed in and vertex data is communicated to other partitions in the form of messages~(fig. \ref{fig:pcpm_model}). 
\item	Intra-partition vertex data updates~(\textit{Gather}) $\rightarrow$  In this phase, incoming messages of a partition are streamed in and the vertex data is updated as per a user-defined function. This phase also generates a list of vertices that become active and will communicate in the next iteration.
\end{enumerate}


GPOP processes multiple partitions in parallel during both phases.
For Scatter, each partition is provided a distinct memory space (called $bin$) to write its outgoing messages. During Gather, a thread exclusively processes all the messages destined to the partitions allocated to it. Thus, GPOP can communicate and compute vertex data without using synchronization primitives - locks or atomics. 

Each thread in GPOP processes one partition at a time. Within each partition (in either phase), vertex data is reused while edges and messages are only accessed once. GPOP ensures high {\it temporal locality} since a partition's vertex data is designed to fit in the private cache of a core. At the same time, GPOP enjoys high {\it spatial locality} of message reads and writes by storing them in consecutive locations in bins. All threads in GPOP synchronize at the end of each Scatter and Gather phase to ensure that all partitions have received and processed all incoming messages before they get overwritten in the next iteration. Algorithm \ref{alg:nibble} describes the execution flow for an iteration of a typical graph application in GPOP.

GPOP chooses between two different scatter modes based on an analytical model that evaluates the tradeoff in work done per active edge versus maximizing main memory performance. In particular, GPOP achieves completely sequential DRAM accesses for scattering partitions with large number of active edges while slightly increasing the work. However, in all cases work done is always within a constant predetermined factor of the number of active edges, thus guaranteeing theoretical efficiency.

GPOP further employs an interleaved Scatter-Gather strategy that allows updates to propagate asynchronously within
an iteration, leading to faster convergence in certain algorithms. The interleaving of Scatter and Gather phases becomes
feasible because tasks in both phases are performed at the granularity of partitions with a single thread responsible for any 
given partition.

\subsection{GPOP System Optimizations}\label{sec:sysOpt}
GPOP decreases DRAM communication volume, improves cache-efficiency and spatial locality while ensuring theoretical efficiency of the underlying implementation through careful system optimizations, as detailed below.

\subsubsection{Graph Partitioning:}\label{sec:partition}
GPOP uses a lightweight index scheme for partitioning vertices into $k$ disjoint equisized sets\footnote{Graph partitioning algorithms such as~\cite{edgePartition1, edgePartition2} can optimize neighbor packing into few partitions, further reducing the number of messages exchanged in GPOP. However there is a trade-off between pre-processing time and the speedup obtained. To demonstrate the fundamental performance improvements delivered by GPOP, we downplay partitioning effects by using only the lightweight scheme described.}. Vertex set $V^p$ of partition $p$ consists of all vertices with indices $\in [p\frac{V}{k}, (p+1)\frac{V}{k})$ while ($E_a^p$) $E^p$  denotes the set of (active) out-edges from $V^p$. $k$ is chosen to be the smallest number satisfying {\it both} the following performance metrics  1) {\it Cache-efficiency}: We make partitions cacheable - the data for $q=V/k$ vertices should fit in the largest private cache~(L2 cache in Xeon servers). 2) {\it Load balance}: Number of partitions should be sufficient to assist dynamic load balancing and avoid threads being idle for long periods of time. For this purpose, we set $k\geq 4t$, where $t$ is the number of threads. 

\subsubsection{Message Generation and Communication:} \label{sec:message}
GPOP drastically reduces total communication volume during Scatter by aggregating messages. (Active) vertex $v$ generates ${}^P\!d_v$ messages (where ${}^P\!d_v$ denotes its out-partition-degree) - exactly one message for each of its neighboring {\it partitions\ } (containing out-neighbors of $v$). We use $r^p = (\sum_{i \in p} {}^P\!d_v)/ E^p$ to denote the average aggregation factor of partition $p$. A message consists of source vertex data and list of all out-neighbors in the neighboring partition. For example, in BFS a vertex can send its visited status or its index~(to compute parents in BFS tree) as the vertex data. These messages are stored  within distinct memory spaces called \textit{bins}, allotted to each partition-pair. Specifically, $bin[p][q]$ is used to store all messages from partition $p$ to partition $q$ within two arrays - \textit{data\_bin[]} and \textit{id\_bin[]} - for storing vertex data and list of out-neighbors, respectively. \\

\begin{algorithm}
	\caption{An Iteration of a Graph algorithm in GPOP}
	\label{alg:nibble}
	\begin{algorithmic}[1]
		\Statex{${}^P\!N_o(v)\rightarrow$ set of neighboring partitions of $v$}
        	\ParFor {$p \in sPartList$} \Comment{\textbf{Scatter}}
			\ForAll {$v\in V^p_a$}
				\ForAll {$q \in N_o^p[v]$}
					\State{\textbf{insert} $  \{ val[v],V^q \cap N_o(v)\}$ in $bin[p][q] $}

				\EndFor	
			\EndFor

			\ForAll {$v\in V^p_a$} \Comment{\textbf{reInit Frontier}}
				\State{$val[v] = update1(val[v])$}
				\If {$^^21cond1(val[v])$} \ \  {$V^p_a = V^p_a \backslash v$}
				\EndIf
			\EndFor
		\EndParFor
		
		\State{$\_\_synchronize()\_\_$}
		
		\ParFor {$p \in gPartList$} \Comment{\textbf{Gather}}

		\ForAll {$msg \in bin[:][p]$} 
		\ForAll {$v\in msg.id$}
		\State	{$val[v] = acc(msg.val, val[v])$}
		\State	{$V^p_a= V^p_a \cup v$}
		\EndFor
		\EndFor
		
		\ForAll {$v\in V^p_a$} \Comment{\textbf{filter Frontier}}
		\State {$val[v] = update2(val[v])$}
        \If {$^^21cond2(val[v])$} \ \ { $V^p_a = V^p_a \backslash v$}
		\EndIf
		\EndFor
		
		\EndParFor
		
		\State{$\_\_synchronize()\_\_$}
		
	\end{algorithmic}
\end{algorithm}

Each partition $p$ is scattered by one thread (that has exclusive access to the corresponding row of bins~$bin[p][:]$), using one of the two communication modes:  \\ \\
\noindent \textbf{Source-centric~(SC) mode:} 
When the number of active edges are small, the thread reads and processes only active vertices in $p$
(lines 2-4, Alg. ~\ref{alg:nibble}). 1)  (Aggregated) messages  from a given active vertex are written 
(in order of destination partitions) 
into the corresponding bins before processing the next vertex. 2) Successive messages to any partition $q$ from active vertices in $p$ are written to contiguous addresses in $bin[p][q]$, labeled as {\it insertion points} of bins. The small number of partitions ensures that {\it all} insertion points used by a thread
are likely to fit in the cache, thus enabling efficient data packing in the cache lines with increased spatial locality.

SC mode is optimal in terms of work done ($\approx r^pE_a^p$ messages in expectation), however, a thread will switch partitions (bins) being written into which can negatively impact the sustained memory bandwidth and overall performance. This motivates an alternate communication mode for GPOP. \\ \\
\noindent\textbf{Partition-centric~(PC) mode:} All vertices in $V^p$ are scattered. To ensure communication-efficiency, (aggregated) messages from $V^p$ are scattered in partition-centric order, i.e messages are {\it not} interleaved between partitions  
(lines 2-4, Alg.~\ref{alg:pcpmDC}). Here we use the Partition-Node bipartite Graph~(PNG) data layout (proposed in~\cite{pcpm}) to Scatter. Using this layout, a thread generates and writes all messages to $bin[p][q]$ before moving to $bin[p][q \!+\! 1]$, thereby ensuring completely sequential DRAM accesses and high sustained memory bandwidth. 

PC mode requires all out-edges of a partition to be traversed. Consequently, the order in which messages are written  always stays the same and adjacency information written once can be \textit{reused} across multiple iterations. Thus, messages to $bin[p][q]$ only need to contain vertex data from $p$, while neighbor identifiers are present in a separate $id\_bin[]$ are written only once during pre-processing, thereby dramatically reducing DRAM communication volume. 

\begin{algorithm}
	\caption{GPOP Scatter in Partition-Centric mode}
	\label{alg:pcpmDC}
	\begin{algorithmic}[1]
        \Statex{$N_i^p(q)\rightarrow$ vertices in $p$ that have out-neighbors in $q$
			}
		\ParFor {$p \in sPartList$} \Comment{\textbf{Scatter}}
				\ForAll {$q\in P$}
					\ForAll {$v\in N_i^p(q)$}
						\State{\textbf{insert} $val[v]$ in $bin[p][q]$}
					\EndFor
				\EndFor
			\EndParFor
	\end{algorithmic}
\end{algorithm} 





The PC and SC modes trade off DRAM bandwidth efficiency for slight increase in work (redundant messages from inactive vertices). To achieve the best of both worlds, GPOP analytically evaluates 
the processing cost 
and chooses the lower cost mode {\it individually} for each partition in {\it every} iteration~(Sec.\ref{sec:commModel}). 

Finally, during Gather, the thread for partition $q$ \textit{exclusively} reads all messages from $bin[:][q]$. These reads generate completely sequential memory accesses (in both modes). All messages in $bin[:][q]$ update vertex data of partition $q$~(which is cached) providing good temporal locality. Thus, GPOP enables \textit{atomic} and \textit{lock-free} communication and computation in both phases with improved cache performance. Many applications also require 
\textit{conditional} reinitialization of  active frontier 
in 
each iteration (parallel Nibble, 
Heat Kernel PageRank \cite{nibble} etc.). GPOP intrinsically supports this functionality. After processing messages in either phase, user functions update (cached) data of active vertices and re-evaluate their active status (lines 5-7, 14-16 in alg.~\ref{alg:nibble}). 



\subsubsection{2-level Active List:}
If Gather is unaware of the bins that are empty~(not written into at all in the preceding Scatter phase), it will probe all the bins performing at least $\theta(k^2)$ work in the process. Theoretically, probing all partitions can make the implementation inefficient since $k=\theta(V)$ (partition size is capped by cache size). Practically, even if $k$ is small~(in thousands for very large graphs), $k^2$ can be large~(in millions). This can heavily deteriorate  performance, especially when the useful work done per iteration could be very small, for instance, in the Nibble algorithm~\cite{spielman}. 

We propose a hierarchical list data structure to avoid these performance inefficiencies. At the top level, we implement a global list, called $gPartList$, that enumerates all partitions~(columns of bins) which have received at least one incoming message in the current iteration. At the second level, we implement a local list for each partition, called $binPartList$, that enumerates all the bins of a given partition which have received at least one incoming message. 
Thus, when a partition $p$ sends a message to partition $q$ during scatter phase, it inserts
\begin{enumerate*}[label={(\arabic*)}]
	\item the message into $bin[p][q]$,
	\item $q$ into $gPartList$ and
	\item $p$ into $binPartList[q]$.
\end{enumerate*}

During Gather, the $gPartList$ serves as a queue of \textit{tasks} to be executed in parallel. The thread allocated to partition $q$ processes all $bin[:][q]$ listed in $binPartList[q]$. Similarly, the Gather phase populates another list enumerating partitions with active vertices to be scattered in the next iteration, and creates a frontier $V^p_a$ of active vertices in partition $p$.


\subsection{Performance Modeling}\label{sec:commModel}
Next we describe GPOP's analytical performance model for choosing the optimal communication mode for each partition.
Let $s_i$ denote the size of a vertex index and $s_v=sizeof(val(v))$ denote  the size of a message data element (vertex value), respectively (both assumed 4B for the applications in this paper). 
Then the total communication volume generated by partition $p$ in SC mode is given by:
\begin{align} 
(V^p_a + E^p_a)s_i + 2(r^pE^p_a s_v + E^p_a s_i) 
\notag
\end{align}
The first term represents Scatter phase traffic for reading $V^p_a$ offsets (CSR) and $E^p_a$ edges. These generate messages containing an expected $r^pE^p_a$ data values and $E^p_a$ destination IDs (second term) -
which are exactly read by the succeeding Gather phase (factor of $2$). Note that it is possible to exactly count messages in SC but we use $r^p$ as a faster approximation. 


The PC mode reads $r^p E^p$ edges and $k$ offsets from the PNG layout and writes $r^p E^p$ messages containing only vertex data. The  Gather phase, however, reads both the vertex data and the destination IDs from these messages. Total DRAM communication for PC mode is given by:
\begin{align}\label{eq:commPC}
r^p E^p s_i + k s_i + 2 r E^p s_v + E^p s_i 
= E^p((r^p + 1)s_i + 2 r s_v) + k s_i 
\notag
\end{align}

Since graph algorithms are mostly memory-bound, we compute the execution time as a ratio of communication volume and bandwidth~$\left(BW_{SC} \text{ or } BW_{PC}\right)$. GPOP chooses PC mode for scattering a partition only if:
\begin{equation}\label{eq:commMode}
	\frac{E^p((r^p+1)s_i + 2rs_v) + ks_i}{BW_{PC}} \leq \frac{2r^p E^p_a s_v + 3 E^p_a s_i + V^p_a s_i}{BW_{SC}}
\end{equation}
 
The ratio $\frac{BW_{PC}}{BW_{SC}}$ is a user configurable parameter which is set to $2$ by default. Note that although PC mode generates more messages, it is selected only if it does work 
within a constant factor of optimal
.
Also note that 
$V^p_a$ ($E^p_a$) for a partition can increase or decrease each iteration (depending on the algorithm). By carefully evaluating the communication mode {\it individually} for each partition in every iteration, GPOP generates work efficient implementations of graph algorithms that achieve high bandwidth for DRAM communication. 

\subsection{Interleaving Scatter and Gather phases}\label{sec:isg}
When scatter and gather phases are completely separated, any update to a vertex is made visible to its neighbors only 
in the next iteration. This can negatively affect the convergence rate for algorithms such as single source shortest path, connected components etc. To prevent such side effects of the two phased Scatter-Gather programming model, GPOP provides a unique \textit{Interleaved Scatter-Gather (ISG)} feature that propagates vertex data updates asynchronously within each iteration, thereby increasing the rate of convergence and reducing overall executiong time of an algorithm.

When ISG is enabled, the Scatter phase proceeds as follows : When a thread is allocated a partition $p$ to scatter, before generating the outgoing messages to its neighboring partitions, the thread first gathers all the incoming messages to $p$ that
have already been received from its in-neighbors. These messages are then processed to update the vertices in $p$ and a new frontier is generated as previously inactive vertices can be activated. Thus, while scattering, updated values of active
vertices in $p$ (instead of the stale values present at the end of the previous iteration) are packed into the messages. 

This interleaving of gather operations within the Scatter phase is made possible by 
\begin{enumerate}[leftmargin=*]
    \item Exclusive thread ownership of partition-level granularity tasks in GPOP - a single thread scatters all the active vertices in a partition $p$ at a time. Therefore, it can also process the messages received by $p$ just before scattering it, as both the tasks operate on the vertex data of the same partition which is loaded into the cache once. Thus, using ISG feature does not affect the cache efficiency of GPOP. 
    
    \item Shared-memory model of target Multicore platforms - ensures that messages written in a bin are immediately visible to the destination partition without requiring the threads to synchronize and explicitly send/receive data. Thus, within the same scatter phase, these messages can be accessed and processed. This is unlike certain distributed-memory models where processes need to synchronize (like threads at the end of Scatter phase) and explicitly exchange messages.
\end{enumerate}

Note that such interleaving of gather operations is not possible in state-of-the-art blocking methods \cite{propagationBlocking, spmv, spmspv}, where incoming messages to a partition's vertices are aggregated together in the same bin, but threads in scatter phase work at vertex-level granularity and different threads may scatter vertices of the same partition at different times.

 \section{GPOP Programming Interface}
\label{sec:fw}
%
%
We carefully assess the program structure in GPOP to construct the right set of APIs that 
\begin{enumerate*}[label={(\alph*)}]
\item hide all the details of parallel programming model from the user and allow an application to be described algorithmically,  
\item do not restrict the functionality of the framework and 
provide enough expressive power to implement a variety of applications. 
\end{enumerate*}

In each iteration, the GPOP back-end calls four main user-defined functions that steer application development:
\begin{enumerate}
\item{\bf \texttt{scatterFunc(node)}}:
Called for all (active) vertices during (SC) Scatter; 
should return an (algorithm-specific) value to propagate (write in bins) to neighbors of \texttt{node} (Eg. alg. \ref{alg:nibblexyz}).
 




\item {\bf \texttt{initFunc(node)}}: Applied to all active vertices at the end of Scatter (Alg. \ref{alg:nibble}: lines 5-7). 
It allows conditional continuation of a vertex in active frontier i.e. 
a vertex \textit{remains} active for the \textit{next} iteration if this function returns \texttt{true}. 
A specific example is parallel Nibble (alg. \ref{alg:nibblexyz}: lines 4-6), where, a vertex stays active if its degree-normalized PageRank is greater than a certain threshold -- $\epsilon$, irrespective of whether it gets updated or not during Gather. The programmer can also use \texttt{initFunc()} to {\it update} vertex data before Gather begins. In Alg. ~\ref{alg:nibblexyz}, for example, only half the $PR$ of a vertex is retained as half is scattered amongst its out-neighbors.


To the best of our knowledge, 
existing frameworks put the onus of implementing such selective continuity 
in the frontier, on the programmer, thus increasing program complexity. 
In contrast, GPOP intrinsically provides such functionality that appreciably reduces programmer effort, 
both in terms of length and complexity of the code.

\item \textbf{\texttt{gatherFunc(val,node)}}: 
For all messages read during Gather phase, this function is called with the data value and destination vertex of the message as input arguments. 
The user can accordingly update the vertex data without worrying about atomicity or locking. 
The vertex under consideration is marked active if the function returns \texttt{true}.


\item \textbf{\texttt{filterFunc(node)}}: This function applies to all active vertices in the preliminary frontier built after message processing and can be used to filter out (deactivate) some vertices.
For example, in parallel Nibble, 
\texttt{filterFunc()} 
performs the threshold check on all active vertices.
The semantics of this function are similar to \texttt{initFunc}. It can also be used to update the final aggregated vertex values computed by \texttt{gatherFunc()}. For instance, in PageRank~(Alg.~\ref{alg:pagerank}), we use this function to apply the damping factor to each vertex.

\end{enumerate}

For the case of weighted graphs, the user can define how the weight should be applied to the source vertex' value, using the \texttt{applyWeight()} function. 



 \section{Applications}\label{sec:app}
In this paper, we evaluate the performance of GPOP on five key applications. 
Algorithms \ref{alg:nibblexyz}-\ref{alg:sssp} illustrate the 
respective user codes in GPOP.
Program initialization and iterative calls to Scatter-Gather execution are similar for all applications and therefore only shown for Alg. \ref{alg:nibblexyz} (lines 13-17). 

\noindent {\bf Parallel Nibble~(Alg. ~\ref{alg:nibblexyz})} -
Computes the probability distribution of a seeded random walk. Nibble is often used for strongly local clustering ~\cite{spielman,nibble}, where it is \textit{mandatory} for the parallel implementation to be work-optimal. 
\begin{algorithm}[htbp]
	\caption{Parallel Nibble implementation in GPOP}
	\label{alg:nibblexyz}
	\begin{algorithmic}[1]
		\State {\textbf{struct $UF$}\{}
		\Procedure{scatterFunc}{node}
		\State {return $PR[\text{node}]/(2*deg[\text{node}]);$
		\EndProcedure
		\Procedure{initFunc}{node}
		\State $PR[\text{node}] = \frac{PR[\text{node}]}{2};\ \ \ \ \ \ \ \ \ \ \ \ \ \ \ \ \ \ \ \ \ \ \ \ \ \ \ \ \ //update1//$
		\State {return $(PR[\text{node}] \geq \epsilon*deg[\text{node}]);\ \ \ \ \ \ \ //cond1//$}
		\EndProcedure
		\Procedure{gatherFunc}{val, node}
		\State {$PR[\text{node}] += \text{val};$}
		\State {}return true;}
		\EndProcedure
		\Procedure{filterFunc}{node}
		\State {return $(PR[\text{node}] \geq \epsilon*deg[\text{node}]);\};\ \ \ \ \ \ \ //cond2//$}
		\EndProcedure

 		\Function{nibble}{$G, v_s$} \Comment{$v_s \rightarrow$ start node}
 		\State{$loadFrontier(G, v_s);$}
 		\State{$PR[] = {0}; PR[v_s] = 1;$}	
 		\While{$(G.frontierSize > 0\ \textbf{and}\ MAX\_ITER)$} 
 		\State{$scatter\_and\_gather(\&G, UF);$} 
 		\EndWhile
 		\EndFunction
	\end{algorithmic}
\end{algorithm}

\noindent {\bf Breadth-First Search~(BFS)~(Alg.~\ref{alg:bfs})} -  Used for rooted graph traversal or search (second kernel in 
Graph500 \cite{graph500}). Algorithm \ref{alg:bfs} finds the parent of every reachable node in the BFS tree rooted at a given vertex.

\begin{algorithm}[htbp]
	\caption{Breadth First Search implementation in GPOP}
	\label{alg:bfs}
	\begin{algorithmic}[1]
		\Statex{ $v_s\rightarrow$ root node;\ $parent[]=\{-1\}; parent[v_s]=v_s$;}
		\Procedure{scatterFunc}{node}
		    \State return node;
		\EndProcedure
		\Procedure{initFunc}{node}
		    \State return false;
		\EndProcedure
		\Procedure{gatherFunc}{val, node}
		    \If {$parent[node]<0$}
		        \State $parent[node]$ = val;
		        \State return true;
		    \EndIf
		    \State return false;
		\EndProcedure
		\Procedure{filterFunc}{node}
		    \State return true;
		\EndProcedure
		
	\end{algorithmic}
\end{algorithm}

\noindent{\bf PageRank~(Alg.~\ref{alg:pagerank})} -
A node ranking algorithm; representative of the crucial SpMV multiplication kernel which is used in many scientific and engineering applications~\cite{parCompLandscape, oski, taoPar}.
We implement the topological PageRank in which all vertices are always active. Hence, \texttt{initFunc} and \texttt{filterFunc} are simply used to reinitialize PageRank values and apply damping factor, respectively.

\begin{algorithm}[htbp]
	\caption{PageRank implementation in GPOP}
	\label{alg:pagerank}
	\begin{algorithmic}[1]
		\Statex{$pageRank[]=\{1/|V|\}$}
		\Procedure{scatterFunc}{node}
		\State return $pageRank[\text{node}]/deg[\text{node}]$;

		\EndProcedure
		
		\Procedure{initFunc}{node}
		\State $pageRank[\text{node}]=0$;
		\State return true;
		\EndProcedure
		
		\Procedure{gatherFunc}{val, node}
		\State $pageRank[\text{node}]$ += val;
		\State return true;
		\EndProcedure
		
		\Procedure{filterFunc}{node}
		\State {$pageRank[\text{node}] = (1-d)/|V| + d*pageRank[\text{node}]$ };
		\State return true;
		\EndProcedure
		
	\end{algorithmic}
\end{algorithm}

		
		

\noindent{\bf Single Source Shortest Path~(SSSP)~(Alg.~\ref{alg:sssp})} - Finds the shortest distance to all vertices in a weighted graph from a given source vertex (third kernel in Graph500). We use Bellman Ford to compute shortest path distances of reachable vertices from a given root vertex.
SSSP uses an additional \texttt{applyWeight()} function to specify that the sum of edge weight~(\texttt{wt}) and source node distance (\texttt{val}) should be propagated to the neighbors.

\begin{algorithm}[htbp]
	\caption{Bellman Ford (SSSP) implementation in GPOP}
	\label{alg:sssp}
	\begin{algorithmic}[1]
		\Statex{$distance[]=\{\infty\}; distance[v_s]=0;$}
		\Procedure{scatterFunc}{node}
		\State return $distance[\text{node}]$;
		\EndProcedure		
		\Procedure{initFunc}{node}
		\State return false;
		\EndProcedure
		\Procedure{gatherFunc}{val, node}
		\If {val $< distance[\text{node}]$}
		\State $distance[\text{node}]$ = val;
		\State return true;
		\EndIf
		\State return false;
		\EndProcedure		
		\Procedure{filterFunc}{node}
		\State return true;
		\EndProcedure
		\Procedure{applyWeight}{val, wt}
		\State return (val + wt);
		\EndProcedure
		
	\end{algorithmic}
\end{algorithm}
%
\noindent{\bf Connected Components (CC)} - We use Label Propagation to compute connected components. All vertices are initialized with their own IDs as the labels. Every active vertex scatters its label to its neighbors and selects the minimum of its current label and the values in incoming messages as its future label. If the label of a vertex changes, it becomes active in the next iteration. The program structure is similar to SSSP as minimum label in neighborhood is selected, except there are no edge weights.

 \section{Evaluation}\label{sec:expEval}
\subsection{Experimental Setup}\label{sec:expSetup}
We conduct experiments on the following 2 machines
\begin{enumerate}[leftmargin=*]
	\item $M1$:  a dual-socket Broadwell server with two 18-core Intel Xeon E5-2695 v4 processors running Ubuntu 16.04.
 	\item $M2$: a dual-socket Ivybridge server with two 8-core Intel Xeon E5-2650 v2 processors running Ubuntu 14.04.
\end{enumerate}
The important properties of these machines are listed in table~\ref{table:sysChars}. Memory bandwidth is measured using STREAM benchmark~\cite{stream}. 

\begin{table}[htbp]
	\centering
	\caption{System Characteristics}
	\label{table:sysChars}
	\begin{tabular}{cc|c|c|}
		\cline{3-4}
		&           & $M1$      & $M2$      \\ \hline
		\multicolumn{1}{|c|}{\multirow{1}{*}{\textbf{Socket}}}             
		& L3 cache  & 45 MB     & 25 MB     \\ \hline
		\multicolumn{1}{|c|}{\multirow{2}{*}{\textbf{Core}}}                                                   & L1d cache & 32 KB     & 32 KB     \\ \cline{2-4} 
		\multicolumn{1}{|c|}{}                                                                                 & L2 cache  & 256 KB    & 256 KB    \\ \hline
		\multicolumn{1}{|c|}{\multirow{3}{*}{\textbf{\begin{tabular}[c]{@{}c@{}}Main \\ Memory\end{tabular}}}} & size      & 1 TB      & 128 GB    \\ \cline{2-4} 
		\multicolumn{1}{|c|}{}                                                                                 & Copy BW   & 55.1 GBps & 43.7 GBps \\ \cline{2-4} 
		\multicolumn{1}{|c|}{}                                                                                 & Add BW    & 64.2 GBps & 48.8 GBps \\ \hline
	\end{tabular}
\end{table}

\paragraph{Implementation Details:} We compile GPOP using G++ 6.3.0 and parallelize it using OpenMP version 4.5. Partition size in GPOP is set to $256$ KB, same as L2 cache size, which has been shown to be a good heuristic \cite{pcpm}. We enable the Interleaved Scatter-Gather (ISG) feature (section \ref{sec:isg} for Connected Components and Single Source Shortest Path computation to accelerate convergence; remaining applications are run in a bulk-synchronous manner. For PageRank, we record our observations over 10 iterations of the algorithm and not until convergence. For the Nibble algorithm, we set the threshold $\epsilon$ to $10^{-9}$. Unless explicitly stated otherwise, the results depict performance using 36 threads on $M1$ and 16 threads on $M2$. 

\paragraph{Datasets:} We use large real world and synthetic graph datasets to evaluate the performance of GPOP. Table~\ref{table:datasets} lists the real-world datasets used in the experiments. \textit{soclj}, \textit{gplus}, \textit{twitter} and \textit{friend} are social networks~; \textit{pld}, \textit{sd1} and \textit{uk} are hyperlink graphs extracted by web crawlers. The synthetic graphs used in this paper are generated by RMAT graph generator with default settings~(scale-free graphs) and degree $16$. All the social networks are small-world graphs with very low diameter ($<40$); the hyperlink graphs have relatively much larger diameter ($>125$ for \textit{uk}). As a result, most of the algorithms converge within few iterations on social networks compared to several hundred iterations on \textit{pld}, \textit{sd1} and \textit{uk}.

The \textit{uk} graph is heavily pre-processed by Webgraph~\cite{webgraph} and LLP~\cite{llp} tools, and has very high locality due to optimized vertex ordering. In order to understand the effects of vertex ordering induced locality on performance of different frameworks, we also create a \textit{uk\_rnd} graph by randomly shuffling the vertex IDs in \textit{uk}. We create weighted graphs for evaluating SSSP by imparting each edge, an integral weight in the range $[1,\log |V|]$. This edge weight distribution has been used in several existing works \cite{ligra, julienne}.
\begin{table}[htbp]
	\centering
	\caption{Graph Datasets}
	\label{table:datasets}
		\begin{tabular}{|c|c|c|c|}
			\hline
			\textbf{Dataset} & \textbf{Description} & \textbf{\# Vertices} & \textbf{\# Edges} \\ \hline
			soclj~\cite{koblenz}           & LiveJournal (social network)      & 4.84 M                   &  68.99 M    \\ \hline
			gplus~\cite{gplus}           & Google+ (social network)         & 28.94 M                   & 462.99 M \\ \hline
			pld~\cite{domain}              & Pay-Level-Domain (hyperlink graph) & 42.89 M                   & 623.06 M  \\ \hline
			twitter~\cite{koblenz, twitter}          & follower network (social network)     & 61.58 M                   & 1468.36 M             \\ \hline
			sd1~\cite{domain}              & Subdomain graph (hyperlink graph)     & 94.95 M                    & 1937.49 M \\ \hline
            friend~\cite{koblenz}              & Friendster (social network)        & 68.35 M                   & 2568.14 M         \\ \hline
            uk-2007~\cite{uk} & UK Domain graph (hyperlink graph) & 105.9 M & 3738.7 M \\ \hline
			rmat\textless$n$\textgreater~\cite{rmat}          & synthetic graph     & $2^n$ M                   & $16*2^n$ M              \\ \hline
		\end{tabular}
\end{table}

\subsection{Results}

\subsubsection{Comparison with Baselines}\label{sec:baselines}
We compare the performance of GPOP against 3 most popular
state-of-the-art 
frameworks - Ligra~\cite{ligra}, GraphMat~\cite{graphmat} and Galois~\cite{galois}.
For BFS, we also report the time for Ligra without direction optimization (Ligra\_Push) for effective comparison.
Galois provides different algorithms for shortest path computation - we report the time for both topological (similar to Bellman Ford) and delta-stepping implementations. Similarly, for Galois Pagerank, we measure the execution time of both residual \cite{galoisPagerank} and the topological versions (recommended pull-direction variants).
GraphMat only supports graphs with $<2^{31}$ edges and cannot process the \textit{friend} and \textit{uk} graphs.
For the parallel Nibble algorithm, the GraphMat and Galois implementations are not available and we only compare against Ligra implementation given in~\cite{nibble}. 

\begin{table}[]
\centering
\caption{Comparison of Execution Time for various Graph algorithms. Red color highlights the best performing framework.}
\resizebox{\linewidth}{!}{
\label{table:baseline}
\begin{tabular}{|c|c|c|c|c|c|c|c|c|c|}
\hline
                                                                                                  &                                      & \multicolumn{8}{c|}{\textbf{Execution time on Datasets (in seconds)}}                                                                                                                                                                                                                                                    \\ \cline{3-10} 
\multirow{-2}{*}{\textbf{Application}}                                                               & \multirow{-2}{*}{\textbf{Framework}} & \textbf{soclj}                       & \textbf{gplus}                       & \textbf{pld}                         & \textbf{twitter}                     & \textbf{sd1}                         & \textbf{friend}                      & \textbf{uk}                            & \textbf{uk\_rnd}                      \\ \hline \hline
                                                                                                  & GPOP                                 & {\color[HTML]{CB0000} \textbf{0.26}} & {\color[HTML]{CB0000} \textbf{1.95}} & {\color[HTML]{CB0000} \textbf{2.5}}  & {\color[HTML]{CB0000} \textbf{4.33}} & {\color[HTML]{CB0000} \textbf{7.17}} & {\color[HTML]{CB0000} \textbf{8.05}} & 4.69                                   & {\color[HTML]{CB0000} \textbf{13.63}} \\ \cline{2-10} 
                                                                                                  & Ligra                                & 1.55                                 & 24.2                                 & 10.7                                 & 29.9                                 & 33.1                                 & 155                                  & 9.59                                   & 80                                    \\ \cline{2-10} 
                                                                                                  & GraphMat                             & 0.31                                 & 4.51                                 & 5.23                                 & 10.1                                 & 15.7                                 & -                                    & -                                      & -                                     \\ \cline{2-10} 
                                                                                                  & Galois(Topological)                  & 0.26                                 & 3.2                                  & 4.5                                  & 8.2                                  & 12.4                                 & 33                                   & {\color[HTML]{CB0000} \textbf{3.42}}   & 21.1                                  \\ \cline{2-10} 
\multirow{-5}{*}{PageRank}                                                                        & Galois(Residual)                     & 0.25                                 & 2.5                                  & 3.66                                 & 7.4                                  & 10.7                                 & 28.7                                 & 3.9                                    & 19.9                                  \\ \hline \hline
                                                                                                  & GPOP                                 & 0.04                                 & 0.37                                 & 0.51                                 & 0.8                                  & 2                                    & 1.31                                 & 1.89                                   & 6.08                                  \\ \cline{2-10} 
                                                                                                  & Ligra                                & {\color[HTML]{CB0000} \textbf{0.04}} & {\color[HTML]{CB0000} \textbf{0.27}} & {\color[HTML]{CB0000} \textbf{0.32}} & {\color[HTML]{CB0000} \textbf{0.49}} & 1.9                                  & {\color[HTML]{CB0000} \textbf{1.25}} & 1.56                                   & 6.33                                  \\ \cline{2-10} 
                                                                                                  & Ligra\_Push                    & 0.07                                 & 0.53                                 & 1.07                                 & 2.81                                 & 4.1                                  & 4.02                                 & 2.1                                    & 3.21                                  \\ \cline{2-10} 
                                                                                                  & GraphMat                             & 0.11                                 & 1.35                                 & 2.44                                 & 2.71                                 & 12.2                                 & -                                    & -                                      & -                                     \\ \cline{2-10} 
\multirow{-5}{*}{BFS}                                                                             & Galois                               & 0.04                                 & 0.32                                 & 0.49                                 & 1.05                                 & {\color[HTML]{CB0000} \textbf{1.01}} & 3.13                                 & {\color[HTML]{CB0000} \textbf{0.63}}   & {\color[HTML]{CB0000} \textbf{1.16}}  \\ \hline \hline
                                                                                                  & GPOP                                 & {\color[HTML]{CB0000} \textbf{0.16}} & {\color[HTML]{CB0000} \textbf{1.21}} & {\color[HTML]{CB0000} \textbf{1.01}} & {\color[HTML]{CB0000} \textbf{1.48}} & 4.5                                  & {\color[HTML]{CB0000} \textbf{3.54}} & 12.17                                  & 18.28                                 \\ \cline{2-10} 
                                                                                                  & Ligra                                & 0.8                                  & 5.6                                  & 6.15                                 & 9.6                                  & 12.9                                 & 14.8                                 & 16                                     & 26                                    \\ \cline{2-10} 
                                                                                                  & GraphMat                             & 0.18                                 & 1.9                                  & 5.22                                 & 4.3                                  & 22.3                                 & -                                    & -                                      & -                                     \\ \cline{2-10} 
                                                                                                  & Galois (Delta-step)                       & 0.19                                 & 1.28                                 & 1.3                                  & 3                                    & {\color[HTML]{CB0000} \textbf{4.04}} & 5.1                                  & {\color[HTML]{CB0000} \textbf{7.2}}    & {\color[HTML]{CB0000} \textbf{12.16}} \\ \cline{2-10} 
\multirow{-5}{*}{SSSP}                                                                            & Galois(Topological)                        & 0.8                                  & 5.1                                  & 6.75                                 & 7.57                                 & 13.6                                 & 18.9                                 & 16.7                                   & 30.4                                  \\ \hline \hline
                                                                                                  & GPOP                                 & 0.09                                 & 0.45                                 & {\color[HTML]{CB0000} \textbf{0.52}} & {\color[HTML]{CB0000} \textbf{0.55}} & {\color[HTML]{CB0000} \textbf{1.8}}  & {\color[HTML]{CB0000} \textbf{2.16}} & 6.03                                   & 11.3                                  \\ \cline{2-10} 
                                                                                                  & Ligra                                & 0.22                                 & 3.3                                  & 1.7                                  & 4.8                                  & 6.9                                  & 21.7                                 & 2.91                                   & 40.1                                  \\ \cline{2-10} 
                                                                                                  & GraphMat                             & 0.15                                 & 2.94                                 & 3.91                                 & 5.73                                 & 16.7                                 & -                                    & -                                      & -                                     \\ \cline{2-10} 
\multirow{-4}{*}{\begin{tabular}[c]{@{}c@{}}Connected\\ Components\end{tabular}}                  & Galois                               & {\color[HTML]{CB0000} \textbf{0.04}} & {\color[HTML]{CB0000} \textbf{0.12}} & 1.12                                 & 1.75                                 & 3.45                                 & 3.92                                 & {\color[HTML]{CB0000} \textbf{2.3}}    & {\color[HTML]{CB0000} \textbf{9.97}}  \\ \hline \hline
                                                                                                  & GPOP                                 & {\color[HTML]{CB0000} \textbf{0.33}} & {\color[HTML]{CB0000} \textbf{0.19}} & {\color[HTML]{CB0000} \textbf{0.33}} & {\color[HTML]{CB0000} \textbf{1.09}} & {\color[HTML]{CB0000} \textbf{0.47}} & {\color[HTML]{CB0000} \textbf{0.4}}  & {\color[HTML]{CB0000} \textbf{0.0067}} & 0.051                                 \\ \cline{2-10} 
\multirow{-2}{*}{\begin{tabular}[c]{@{}c@{}}Parallel \\ Nibble\end{tabular}} & Ligra                                & 1.6                                  & 0.5                                  & 1.2                                  & 5.3                                  & 1.18                                 & 0.54                                 & 0.0082                                 & {\color[HTML]{CB0000} \textbf{0.022}} \\ \hline
\end{tabular}}
\end{table}


\paragraph{Execution Time:} Table~\ref{table:baseline} provides the runtime of the five graph applications discussed in section~\ref{sec:app}, implemented using GPOP and baseline frameworks. 
For all applications other than BFS, GPOP consistently outperforms the baselines for most datasets with relatively larger speedups for large small-world networks. 

GPOP provides the fastest PageRank implementation for all datasets except \textit{uk}. Specifically, on large datasets \textit{twitter}, \textit{sd1} and \textit{friend}, it achieves up to $19\times$, $2.3\times$ and $3.6\times$ speedup over Ligra, Graphmat and the best performing variant in Galois, respectively. This is because the aggressive cache and memory bandwidth optimizations of GPOP become extremely crucial for performance as the graphs become larger. Further, in PageRank, all vertices are always active allowing GPOP to use the high performance PC mode throughout the course of the algorithm. 

For BFS execution, Ligra is the fastest  for most datasets. The direction optimizing BFS of Ligra drastically reduces the
amount of work done. GPOP does not employ any direction optimization, however, its performance for large graphs \textit{sd1}, \textit{friend} and \textit{uk} is still comparable to that of Ligra.

In case of single source shortest path, GPOP's performance exceeds the other frameworks. It is up to $2\times$ faster than Galois even though Galois utilizes the more efficient (from algorithmic perspective) Delta stepping instead of the Bellman Ford of other frameworks. Compared to GraphMat and Ligra, GPOP is up to $6.5\times$ and $5\times$ faster, respectively. For Connected Components also, GPOP outperforms other frameworks with significantly large speedups for \textit{twitter}, \textit{sd1} and \textit{friend}.
Note that even with ISG enabled, the GPOP typically takes more iterations to converge than Galois or Ligra, because of the larger task granularity. However, the memory access efficiency results in each iteration of GPOP being dramatically faster than the baselines.  

In parallel Nibble, the algorithm explores very few vertices in local neighborhood of the seed node(s), enabling good locality in the compact hash tables used by Ligra. Due to the limited exploration, execution time of Parallel Nibble is not solely affected by the graph size as we see that it runs in few milliseconds for the large \textit{uk} dataset. 
Since all threads are updating a small set of vertices, GPOP benefits by the atomic free updates. Furthermore, the Ligra implementation~\cite{nibble} uses hash-based sparse data structures and suffers from hash collisions. Instead, our implementation uses an array to store probabilities but still achieves theoretical efficiency by amortizing the array initialization costs across multiple runs of Nibble.

The \textit{uk} dataset possesses an optimized vertex ordering that introduces high locality during neighborhood access. Ligra and Galois benefit heavily from this locality resulting in Galois being the best performing framework for \textit{uk}. However, the pre-processing required to generate such a vertex ordering is extremely time consuming and is not always a pragmatic approach. 

\paragraph{Benefits of Vertex Reordering} can be seen by comparing the execution times for \textit{uk} and \textit{uk\_rnd}. The random shuffling disturbs the optimized vertex IDs and reduces locality. In GPOP, using an optimized ordering can reduce the number of messages exchanged between partitions, as neighbors of a vertex are likely to be packed in very few partitions. Hence, we see a $3.2\times$ difference in BFS performance for GPOP. As an example, the number of inter-partition edges in \textit{uk} was $179$ million as opposed to $3.4$ billion for \textit{uk\_rnd}. In other real-world datasets, the number of inter-partition edges were typically half of the total edges in the graph.

However, GPOP's programming model ensures cache efficiency even for \textit{uk\_rnd}. Therefore, the performance difference between \textit{uk} and \textit{uk\_rnd} for Galois and Ligra is much higher than GPOP. 
We also observed that SSSP in Galois converged much faster ($<60$ iterations) compared to GPOP ($>200$ iterations) for both \textit{uk} datasets, which can perhaps be attributed to its priority driven task scheduler.

\begin{table}[]
    \centering
    \caption{Comparison of L2 Cache Misses incurred by GPOP. Ligra's SSSP implementation couldn't feasibly process \textit{uk} graph on $M2$ machine due to large memory requirements for graph IO.}
    \label{table:cache}
    \resizebox{\linewidth}{!}{
\begin{tabular}{|c|c|c|c|c|c|c|c|c|c|}
\hline
                                                                                 &                                      & \multicolumn{8}{c|}{\textbf{L2 Cache misses (in Billions)}}                                                                                                                                                                                                                                                                  \\ \cline{3-10} 
\multirow{-2}{*}{\textbf{Application}}                                           & \multirow{-2}{*}{\textbf{Framework}} & \textbf{soclj}                       & \textbf{gplus}                        & \textbf{pld}                          & \textbf{twitter}                      & \textbf{sd1}                          & \textbf{friend}                       & \textbf{uk}                           & \textbf{uk\_rnd}                      \\ \hline\hline
                                                                                 & GPOP                                 & {\color[HTML]{CB0000} \textbf{0.18}} & {\color[HTML]{CB0000} \textbf{1.44}}  & {\color[HTML]{CB0000} \textbf{1.99}}  & {\color[HTML]{CB0000} \textbf{4.1}}   & {\color[HTML]{CB0000} \textbf{4.95}}  & {\color[HTML]{CB0000} \textbf{7.71}}  & {\color[HTML]{CB0000} \textbf{0.53}}  & {\color[HTML]{CB0000} \textbf{17.28}} \\ \cline{2-10} 
                                                                                 & Ligra                                & {\color[HTML]{000000} 1.06}          & {\color[HTML]{000000} 9}              & {\color[HTML]{000000} 15.18}          & {\color[HTML]{000000} 31.09}          & {\color[HTML]{000000} 47.91}          & {\color[HTML]{000000} 35.18}          & {\color[HTML]{000000} 5.44}           & {\color[HTML]{000000} 135.9}              \\ \cline{2-10} 
                                                                                 & GraphMat                             & {\color[HTML]{000000} 0.33}          & {\color[HTML]{000000} 3.78}           & {\color[HTML]{000000} 6.28}           & {\color[HTML]{000000} 14.92}          & {\color[HTML]{000000} 17.77}          & {\color[HTML]{000000} -}              & {\color[HTML]{000000} -}              & {\color[HTML]{000000} -}              \\ \cline{2-10} 
\multirow{-4}{*}{PageRank}                                                       & Galois                               & {\color[HTML]{000000} 0.39}          & {\color[HTML]{000000} 3.53}           & {\color[HTML]{000000} 5.71}           & {\color[HTML]{000000} 13.25}          & {\color[HTML]{000000} 18.97}          & {\color[HTML]{000000} 42.16}          & {\color[HTML]{000000} 0.9}            & {\color[HTML]{000000} 42.4}           \\ \hline \hline
                                                                                 & GPOP                                 & {\color[HTML]{CB0000} \textbf{0.02}} & {\color[HTML]{CB0000} \textbf{0.13}}  & {\color[HTML]{CB0000} \textbf{0.2}}   & 0.36  & {\color[HTML]{CB0000} \textbf{1.01}}  & {\color[HTML]{CB0000} \textbf{0.51}}  & {\color[HTML]{000000} 0.29}           & {\color[HTML]{000000} 2.84}           \\ \cline{2-10} 
                                                                                 & Ligra                                & {\color[HTML]{000000} 0.03}          & {\color[HTML]{000000} 0.21}           & {\color[HTML]{000000} 0.23}           & {\color[HTML]{CB0000} \textbf{0.34}}           & {\color[HTML]{000000} 2.14}           & {\color[HTML]{000000} 1.2}            & {\color[HTML]{000000} 0.3}           & {\color[HTML]{000000} 13.7}              \\ \cline{2-10} 
                                                                                 & GraphMat                             & {\color[HTML]{000000} 0.03}          & {\color[HTML]{000000} 0.46}           & {\color[HTML]{000000} 0.65}           & {\color[HTML]{000000} 1.35}           & {\color[HTML]{000000} 2.52}           & {\color[HTML]{000000} -}              & {\color[HTML]{000000} -}              & {\color[HTML]{000000} -}              \\ \cline{2-10} 
\multirow{-4}{*}{BFS}                                                            & Galois                               & {\color[HTML]{000000} 0.05}          & {\color[HTML]{000000} 0.39}           & {\color[HTML]{000000} 0.63}           & {\color[HTML]{000000} 1.54}           & {\color[HTML]{000000} 1.69}           & {\color[HTML]{000000} 4.27}           & {\color[HTML]{CB0000} \textbf{0.29}}  & {\color[HTML]{CB0000} \textbf{1.18}}  \\ \hline \hline
                                                                                 & GPOP                                 & {\color[HTML]{000000} 0.14}          & {\color[HTML]{CB0000} \textbf{0.81}}  & {\color[HTML]{CB0000} \textbf{0.87}}  & {\color[HTML]{CB0000} \textbf{1.32}}  & {\color[HTML]{CB0000} \textbf{3.49}}  & {\color[HTML]{CB0000} \textbf{3.58}}  & {\color[HTML]{CB0000} \textbf{2.29}}  & {\color[HTML]{000000} 19.58}          \\ \cline{2-10} 
                                                                                 & Ligra                                & {\color[HTML]{CB0000} \textbf{0.13}} & {\color[HTML]{000000} 0.97}           & {\color[HTML]{000000} 1.48}           & {\color[HTML]{000000} 2.64}           & {\color[HTML]{000000} 4.79}           & {\color[HTML]{000000} 8.13}           & {\color[HTML]{000000} -}              & {\color[HTML]{000000} -}              \\ \cline{2-10} 
                                                                                 & GraphMat                             & {\color[HTML]{000000} 0.14}          & {\color[HTML]{000000} 1.93}           & {\color[HTML]{000000} 3.34}           & {\color[HTML]{000000} 4.57}           & {\color[HTML]{000000} 10.37}          & {\color[HTML]{000000} -}              & {\color[HTML]{000000} -}              & {\color[HTML]{000000} -}              \\ \cline{2-10} 
\multirow{-4}{*}{SSSP}                                                           & Galois                               & {\color[HTML]{000000} 0.18}          & {\color[HTML]{000000} 1.26}           & {\color[HTML]{000000} 1.47}           & {\color[HTML]{000000} 3.4}            & {\color[HTML]{000000} 5.44}           & {\color[HTML]{000000} 6.93}           & {\color[HTML]{000000} 4.37}           & {\color[HTML]{CB0000} \textbf{11.42}} \\ \hline \hline
                                                                                 & GPOP                                 & {\color[HTML]{000000} 0.08}          & {\color[HTML]{CB0000} \textbf{0.3}}   & {\color[HTML]{CB0000} \textbf{0.54}}  & {\color[HTML]{CB0000} \textbf{0.49}}  & {\color[HTML]{CB0000} \textbf{1.53}}  & {\color[HTML]{CB0000} \textbf{1.89}}  & {\color[HTML]{000000} 0.98}           & {\color[HTML]{CB0000} \textbf{8.68}}  \\ \cline{2-10} 
                                                                                 & Ligra                                & {\color[HTML]{000000} 0.18}          & {\color[HTML]{000000} 1.71}           & {\color[HTML]{000000} 3.01}           & {\color[HTML]{000000} 5.11}           & {\color[HTML]{000000} 11.65}          & {\color[HTML]{000000} 19.45}          & {\color[HTML]{000000} 0.54}           & {\color[HTML]{000000} 55.9}              \\ \cline{2-10} 
                                                                                 & GraphMat                             & {\color[HTML]{000000} 0.12}          & {\color[HTML]{000000} 2.17}           & {\color[HTML]{000000} 3.07}           & {\color[HTML]{000000} 4.68}           & {\color[HTML]{000000} 10.42}          & {\color[HTML]{000000} -}              & {\color[HTML]{000000} -}              & {\color[HTML]{000000} -}              \\ \cline{2-10} 
\multirow{-4}{*}{\begin{tabular}[c]{@{}c@{}}Connected\\ Components\end{tabular}} & Galois                               & {\color[HTML]{CB0000} \textbf{0.05}} & {\color[HTML]{000000} 0.5}            & {\color[HTML]{000000} 0.88}           & {\color[HTML]{000000} 1.68}           & {\color[HTML]{000000} 3.16}           & {\color[HTML]{000000} 5.25}           & {\color[HTML]{CB0000} \textbf{0.62}}  & {\color[HTML]{000000} 16.64}          \\ \hline \hline
                                                                                 & GPOP                                 & {\color[HTML]{CB0000} \textbf{0.2}}  & {\color[HTML]{CB0000} \textbf{0.112}} & {\color[HTML]{CB0000} \textbf{0.225}} & {\color[HTML]{CB0000} \textbf{0.705}} & {\color[HTML]{000000} 0.425}          & {\color[HTML]{CB0000} \textbf{0.289}} & {\color[HTML]{CB0000} \textbf{0.001}} & {\color[HTML]{CB0000} \textbf{0.025}} \\ \cline{2-10} 
\multirow{-2}{*}{\begin{tabular}[c]{@{}c@{}}Parallel \\ Nibble\end{tabular}}     & Ligra                                & {\color[HTML]{000000} 0.453}         & {\color[HTML]{000000} 0.288}          & {\color[HTML]{000000} 0.331}          & {\color[HTML]{000000} 4.215}          & {\color[HTML]{CB0000} \textbf{0.386}} & {\color[HTML]{000000} 1.139}          & {\color[HTML]{000000} 0.004}          & {\color[HTML]{000000} 0.012}              \\ \hline
\end{tabular}}
\end{table}

\paragraph{Cache Performance:} We report the L2 cache misses of GPOP and the baselines for the applications used in this paper. For Galois and Ligra, we use the best performing variants for respective applications as noted by their execution time performance.

Because of the aggressive cache optimizations (section~\ref{sec:prog}), GPOP incurs dramatically lesser cache misses than Ligra, GraphMat and Galois for almost all datasets and applications. Even for BFS where Ligra performs much less work than GPOP due to direction optimization, cache misses for GPOP are significantly smaller than Ligra. The benefits of GPOP optimizations are most noticeable for large datasets in PageRank, where all frameworks execute $10$ iterations with a large fraction of vertices (or all vertices for Ligra, GraphMat and GPOP) being active. Consequently, GPOP achieves up to $9\times$, $3.6\times$ and $5.5\times$ lesser cache misses than Ligra, GraphMat and Galois, respectively.

For certain applications on \textit{uk} and \textit{uk\_rand} graphs, Galois converges much faster, thereby reducing the overall memory accesses and consequently, the cache misses as well. The same effect is also seen in Delta stepping based SSSP implementation of Galois, especially for graphs with relatively large diameter such as \textit{sd1}. Cache misses for Parallel Nibble are much smaller as the algorithm only explores a small region around the seed node.

We also observe the impact of vertex ordering induced locality for various frameworks as the cache misses incurred for \textit{uk} increase when the vertex IDs are randomly shuffled (\textit{uk\_rnd}). For PageRank and Connected Components, all of the frameworks show a rapid increase in cache misses but relatively, GPOP remains less affected. We note that both PageRank and Connected 
Surprisingly, the impact of vertex ordering on BFS and SSSP implementations in Galois is significantly smaller than other
applications.

\subsubsection{Scalability}\label{exp:scale}
We evaluate the scalability of GPOP on the target applications using synthetic \textit{rmat} graphs. 
To obtain insights into the performance of GPOP, we need to vary graph parameters, such as \# vertices, average degree etc. in a controllable fashion, which motivates the use of synthetic graphs for this set of experiments.

\begin{figure}[htbp]
	\centering
	\includegraphics[width=1.0\linewidth]{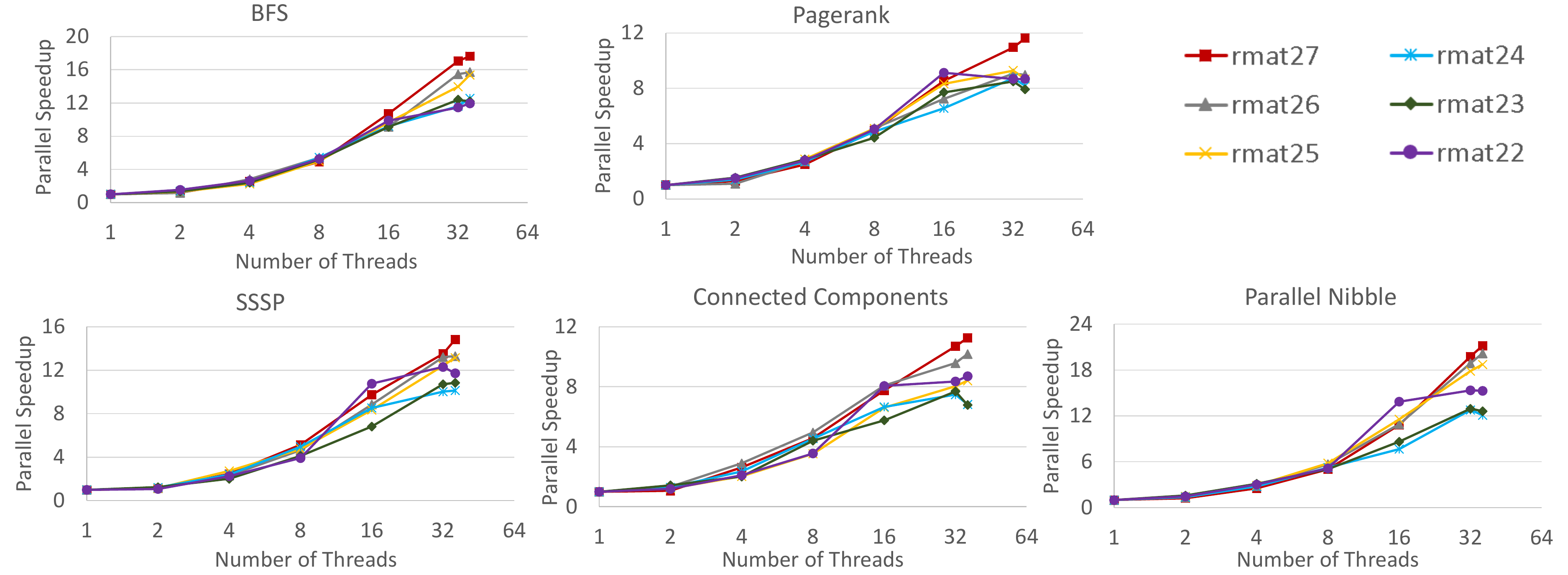}
	\caption{Parallel Speedup (self-relative) of GPOP over single-threaded execution}
	\label{fig:strongscale}
\end{figure}

\paragraph{Strong scaling (figure \ref{fig:strongscale}):}
To analyze strong scaling behaviour of GPOP, we use synthetic graphs with size ranging from \textit{rmat22}$\left(\abs{V}=4M, \abs{E}=64 M\right)$ to \textit{rmat27}$\left(\abs{V}=128M, \abs{E}=2048M\right)$ and record the execution time as the number of threads is varied.

GPOP demonstrates very good scalability for BFS, APPR and SSSP, achieving up to $17.6\times$, $15\times$ and $21.2\times$ parallel speedup using $36$ threads, respectively. Note that the speedup is higher for large datasets that perform more work per iteration reducing the relative overhead of parallelization and synchronization. 

In case of PageRank, GPOP achieves up to $11.6\times$ speedup with 36 threads and scales poorly after 16 threads. This is because PageRank always uses PC mode scatter and nearly saturates the bandwidth with $\approx20$ threads. Therefore, any further increase in number of threads with the same memory bandwidth does not improve overall performance. A similar effect is also seen in Connected Components execution where all vertices are active in the initial iteration and a large fraction of overall execution time is spent in the PC mode. Contrarily, we observe that BFS and SSSP spend relatively more time in SC mode whereas APPR executes entirely in SC mode.

We also observe that performance scaling of GPOP improves as the graph size increases. This is because bigger graphs provide more workload per partition, especially when a large number of threads are used, thereby reducing the relative overheads of parallelization and synchronization. 

\begin{figure}[htbp]
	\centering
	\includegraphics[width=\linewidth]{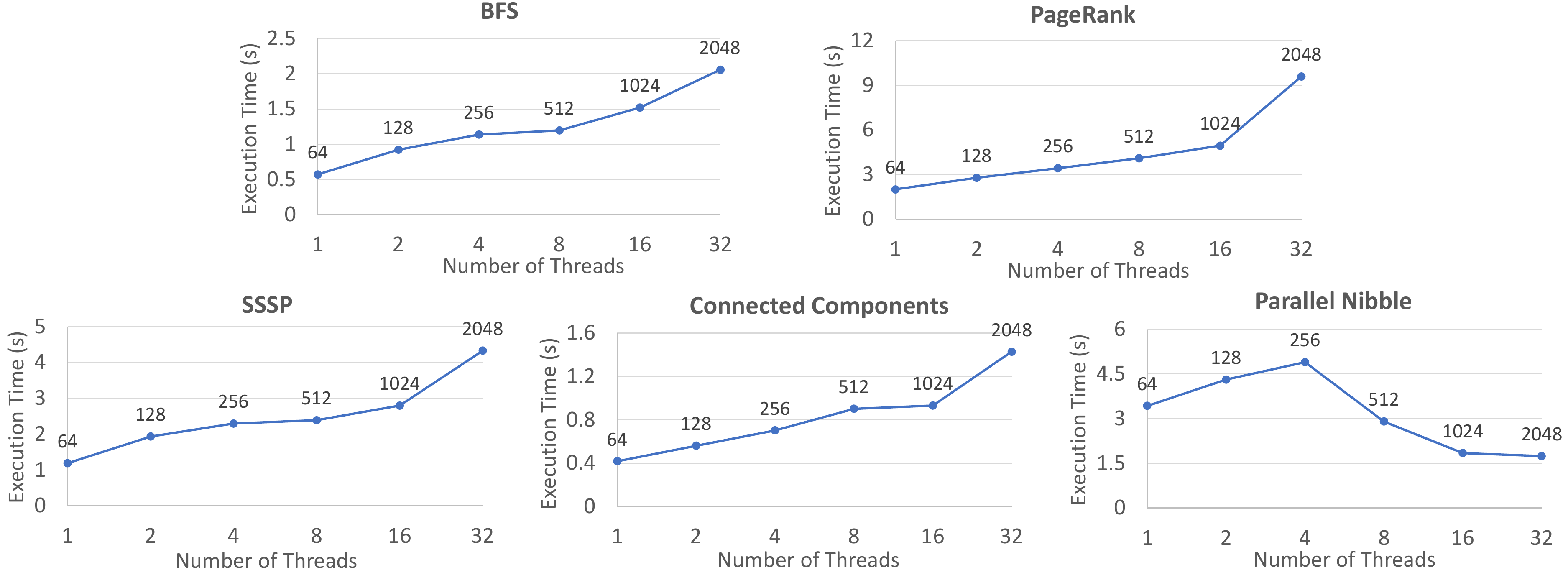}
	\caption{Weak scaling results for GPOP. Labels on the figure indicate \#edges(in Millions)}
	\label{fig:weakscale}
\end{figure}
\paragraph{Weak scaling (figure~\ref{fig:weakscale}):}
To observe the weak scaling behavior of GPOP, we vary the number of threads between $1$ to $32$
along with a proportional change in the size of the input graph being processed. For all of the applications, 
GPOP exhibits less than $4\times$ increase in execution time for a $32\times$ increase in the input graph size.
For PageRank and Connected Components, we see a sharp increase in execution time when going from $16$ to $32$ threads, relative to their remarkable scaling from $1$ to $16$ threads. This can be attributed to the memory bandwidth bottleneck as explained previously.

The behavior of Parallel Nibble is more complex than the other applications. Initially, the time increases with the graph 
size due to an increase in the size of explored subgraph. However, it suddenly drops when going from \textit{rmat24} to \textit{rmat25}. This is because the degree of our seed node increases with graph size, thereby diluting the contribution of the seed to the visiting probability of its neighbors. After a certain point, this contribution becomes too small and is able to activate only a few nodes for the future iterations. Consequently, frontier size and hence, the execution time decreases sharply.


\begin{figure}[htbp]
	\centering
	\includegraphics[width=\linewidth]{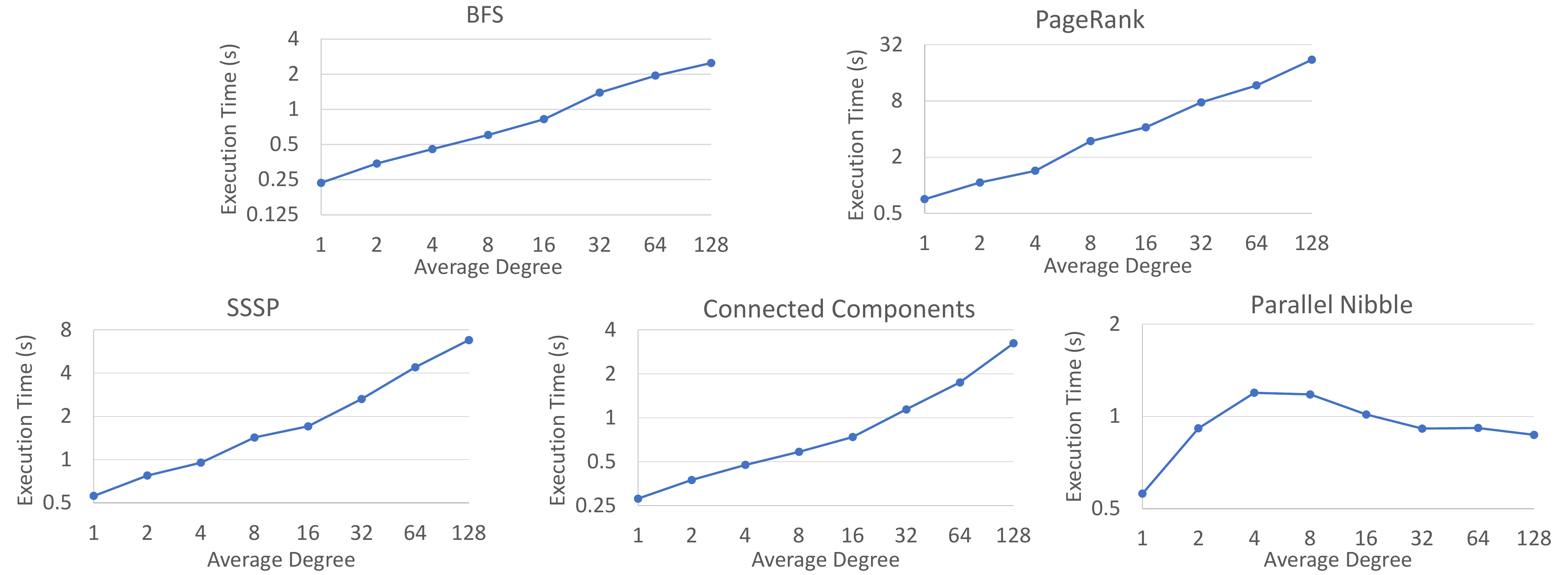}
	\caption{Execution time of GPOP grows linearly with average degree of input graph (\textit{rmat26}, $|V|=64$ M). Both axes in this figure are plotted on logarithmic scale.}
	\label{fig:degreevar}
\end{figure}
\paragraph{Execution Time vs Degree (figure~\ref{fig:degreevar}):}
In order to understand the effects of average degree of the graph on GPOP's performance, we take the \textit{rmat26} ($\abs{V}=64$ million) dataset and vary its degree from $1$ ($\abs{E}=64$ million) to $128$ ($\abs{E}=8$ billion). Figure~\ref{fig:degreevar} clearly shows that the performance of GPOP scales very well with the density of \textit{rmat26}. 

For BFS, SSSP and Connected Components, the execution time increases by less than $12\times$ for a $128\times$ increase in
the average degree. We observed that as the graph became denser, GPOP execution tilted more towards the highly efficient PC mode, which explains the relatively small increase in execution time. This is because the increasing density results in an increase in active frontier size in initial iterations as more vertices are quickly reached. Further, the aggregation factor $r^P$ (section \ref{sec:message}) of partitions increases with degree, making PC mode even more efficient. Contrarily, execution time of PageRank increases by $32\times$ because PageRank always uses PC mode irrespective of the average degree.

Execution time of Parallel Nibble is not determined by the total number of edges in the graph but only by the local region
around the seed node which gets activated. Further, as explained previously (analysis of weak scaling), execution time may decrease as the degree of seed node increases. Hence, we see that the Parallel Nibble execution time does not vary significantly for degree more than $16$.
\begin{figure}[htbp]
	\centering
	\includegraphics[width=\linewidth]{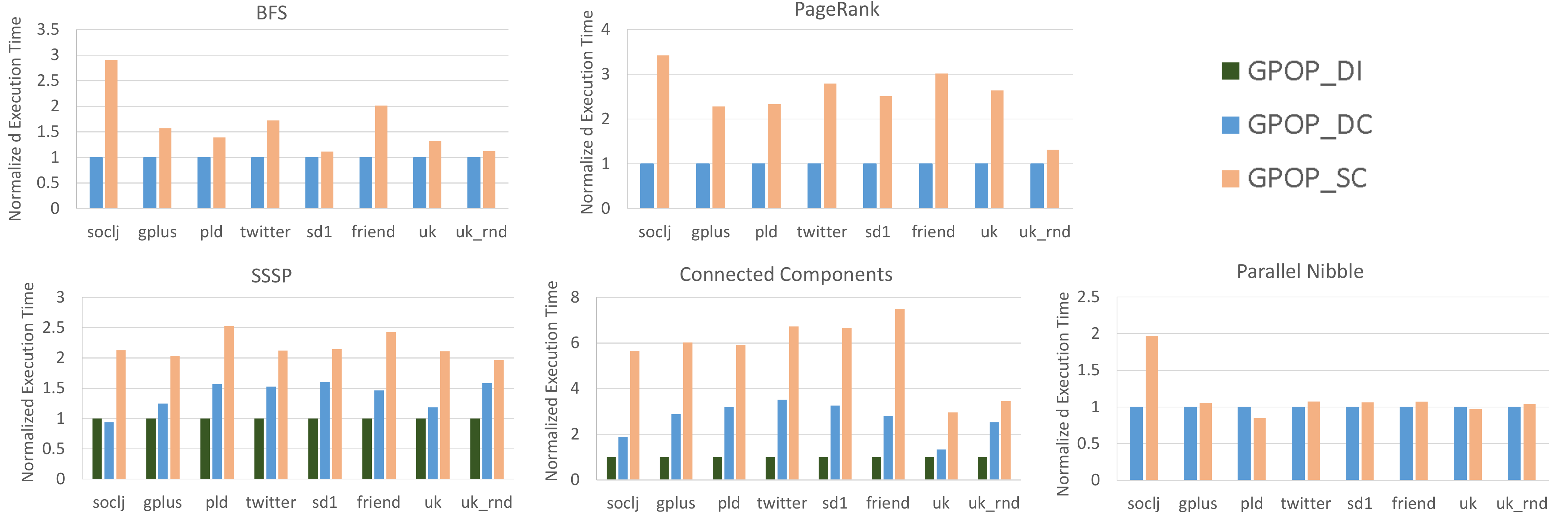}
	\caption{Execution time of GPOP\_SC (only Source-centric communication, no ISG), GPOP\_DC (dual communication mode, no ISG) and GPOP\_DI (dual communication mode with ISG), normalized with GPOP\_DI for SSSP and Connected Components, and GPOP\_DC for remaining algorithms.}
	\label{fig:optimizations}
\end{figure}
\subsubsection{Impact of GPOP Optimizations:} \label{sec:expOpt}
In this subsection, we analyze the impact of the two main optimizations employed by GPOP -- dual-mode communication and Interleaved Scatter-Gather (ISG) operations. Recall that the ISG feature is enabled for SSSP and Connected Components only.

To analyze the individual effect of this optimizations on the overall performance, we measure the execution time for three GPOP variants -- (a) GPOP with only Source-centric communication and ISG disabled (GPOP\_SC), (b) GPOP with dual communication mode and ISG disable (GPOP\_DC), and (c) GPOP with dual communication mode and ISG enabled (GPOP\_DI). 

Figure~\ref{fig:optimizations} shows the relative execution time of GPOP with different optimizations. Clearly, both the optimizations contribute significantly to performance enhancement. Dual communication provides $2\times-3\times$ speedup for PageRank and Connected Components, and $1.5\times-2\times$ speedup for BFS and SSSP. The ISG feature enables faster convergence resulting in a further $3\times$ and $1.5\times$ approx. speedup for Connected Components and SSSP, respectively. 
Note that due to limited graph exploration in Parallel Nibble, the GPOP\_DC is unable to utilize the Partition-centric communication and hence, does not provide any significant speedup over GPOP\_SC.

\paragraph{Communication Mode Comparison:}
\begin{figure}
	\includegraphics[width=\textwidth]{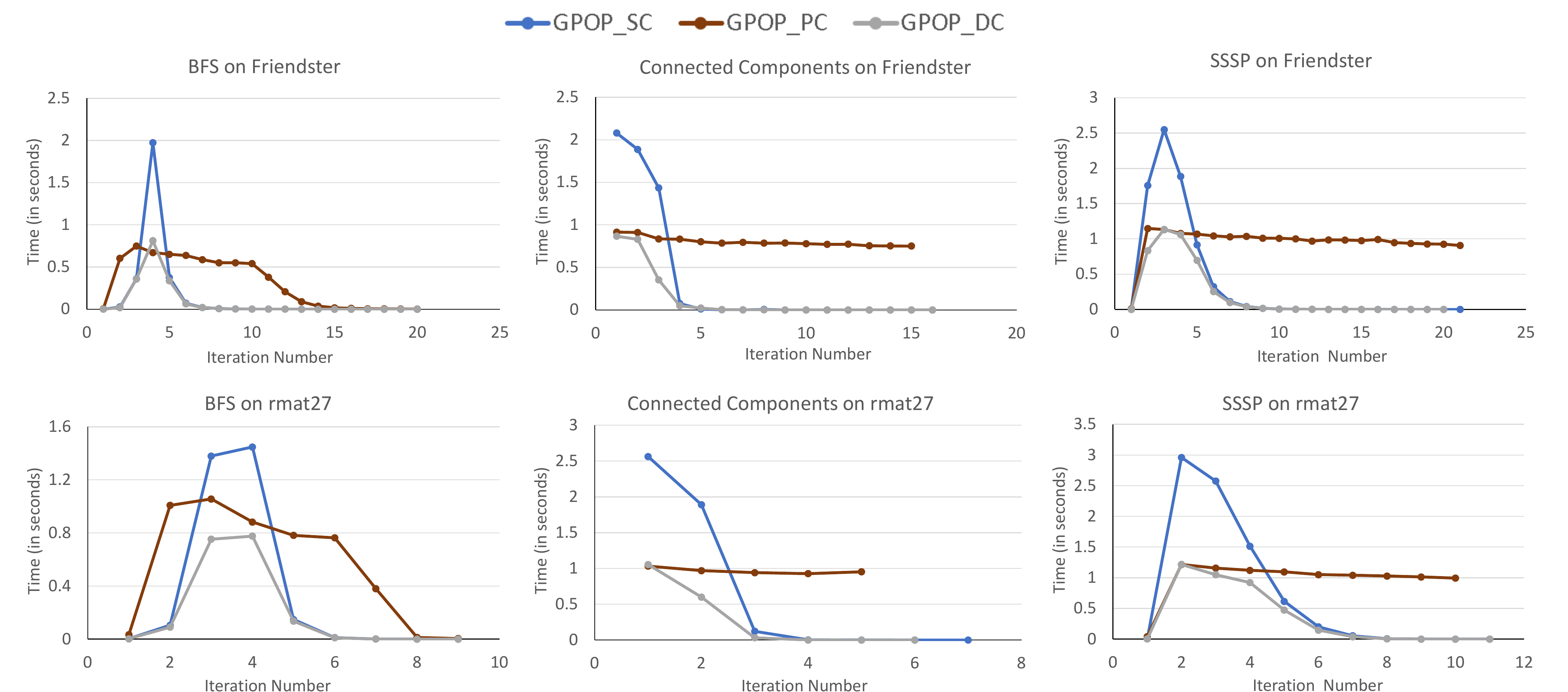}
	\caption{GPOP\_DC combines the best of Source-centric and Partition-centric communication (ISG enabled
	in this experiment for SSSP and Connected Components).}
	\label{fig:spvsden}
\end{figure}
To better understand the behavior of GPOP under different communication modes, we plot per iteration execution time for each of the three modes - GPOP\_SC (Source-centric), GPOP\_PC (Partition-centric) and GPOP\_DC (dual mode).
Note that for this set of experiments, ISG is enabled for SSSP and Connected Components so that we can
observe the differences arising in GPOP's performance when only the scattering approach is varied.

Figure~\ref{fig:spvsden} depicts the execution time of each iteration for BFS, Connected Components and SSSP~(Bellman-Ford), for two large datasets - \textit{friendster} and \textit{rmat$27$}($\abs{V}=128$ million, $\abs{E}=2$ billion). In BFS and SSSP, frontier is dense (lot of active vertices) in the middle iterations and very sparse in the initial and last few iterations. In Connected Components, all vertices are active initially and the frontier size decreases as algorithm proceeds.

Execution time of GPOP\_PC stays constant for almost all but a few iterations. This is because in these iterations, the 2-level hierarchical list prevents scattering of partitions with no active vertices. Runtime of GPOP\_SC varies drastically with frontier size and is larger than GPOP\_PC for iterations with dense frontier. GPOP\_DC achieves the best of both worlds by analytically predicting and choosing the mode with better performance. In almost all iterations, GPOP\_DC executes faster than both GPOP\_SC and GPOP\_PC which empirically validates our analytical model~(section~\ref{sec:commModel}).

We do not show the per iteration time of Pagerank in figure \ref{fig:spvsden} because PageRank is always executed using Partition-centric mode. Moreover, all vertices are active in every iteration and hence, the overall performance difference between GPOP\_DC and GPOP\_SC (figure \ref{fig:optimizations}) also reflects the per iteration performance.
Figure \ref{fig:spvsden} also excludes parallel Nibble as it activates very few vertices in every iteration. Consequently, the dual mode communication does not make any difference and GPOP\_DC is essentially similar to GPOP\_SC.
 \section{Related Work}\label{sec:related}

Pregel~\cite{pregel} was one of the earliest (distributed-memory) graph processing frameworks that proposed the 
Vertex-centric (VC) programming model with the Gather-Apply-Scatter (GAS) abstraction, inspired by the Bulk Synchronous Parallel model. Apache Giraph \cite{facebook} extended this model with added features such as out-of-core computation. However, this model requires synchronizing and explicit message exchange between compute nodes, thus introducing scalability challenges. Some frameworks also support asynchronous execution on distributed-memory systems \cite{powergraph, graphlab} although it sacrifices the efficiency of bulk communication. 

The need to reduce synchronization and communication motivated the development of subgraph-centric frameworks \cite{thinklikeagraph, goffish, blogel}. Such frameworks compute on the local subgraph owned by a machine for several iterations or until local convergence. This is followed by a synchronization step where local updates are sent to other machines and external updates are received, that again kickstart computation on the local subgraph. Contrarily, GPOP is designed for shared-memory systems where messages generated by a subgraph (partition) are immediately visible to other partitions, allowing the use of interleaved scatter and gather. The shared-memory model also allows simple dynamic task allocation in GPOP, where threads are assigned partitions during runtime and are not tied to specific partitions at the beginning, thereby reducing the drawbacks of synchronization. Hence, we only implement one level of intra-partition computation within the ISG feature (section \ref{sec:isg}). 

Many shared memory graph analytic frameworks~\cite{ligra,polymer,grazelle} also adopt the VC model with \textit{push} or \textit{pull} direction execution (algorithm~\ref{alg:pushpull}). The direction optimizing BFS\cite{diroptimizing} performs a frontier dependent switch between the two directions. Ligra~\cite{ligra} is a framework that generalizes this idea to other graph algorithms. 
Galois \cite{galois} utilizes speculative parallelization along with a fine-grained work-stealing based load balancing. Its priority drive task scheduler is very powerful. Especially, for applications like SSSP on large diameter graphs, priority (distance) based exploration of vertices has been shown to significantly reduce the total work done per execution \cite{ssspgpu}. Galois was arguably the fastest baseline in our experiments (section \ref{sec:expEval}).
However, both Galois and Ligra require atomic instructions and are not cache and memory efficient due to low-locality fine-grained random memory accesses. Graph reordering \cite{metis, gorder, laika, webgraph} can improve locality and reduce the synchronization overheads incurred by a scheduler in data graph computations \cite{laika}. However, these methods can be prohibitively expensive for large graphs \cite{slotacase}.


Several works~\cite{xstream,graphmat,spmspv,hpec,spmv} borrow the 2-phased GAS abstraction (also used by GPOP) and avoid use of synchronization primitives on shared-memory platforms. GraphMat \cite{graphmat} maps graph algorithms to efficient Sparse Matrix (SpM) kernels but its bit-vector representation of frontier requires $\mathcal{O}(V)$ work in each iteration. X-Stream uses Edge-centric programming and performs $\mathcal{O}(E)$ traversal in each iteration. Modified EC or blocking techniques \cite{hpec,spmv} have also been utilized to improve locality and cache performance. 
The partitioning based PageRank \cite{pcpm}
sequentializes memory accesses to enhance sustained bandwidth. However, these works~\cite{hpec, spmv, pcpm} are only applicable to Sparse Matrix Dense Vector multiplication
or PageRank computation, where all vertices are active. Very recently, the PartitionedVC framework \cite{matam2019partitionedvc} was proposed for efficient out of core graph processing.
PartitionedVC loads only active vertex adjacencies from the SSD and supports asynchronous processing by updating vertex data before scheduling its adjacency list for processing. 

The benefits of partitioning to improve locality have also been explored previously. However, novelty in the GPOP framework lies in its guaranteed cache performance without optimized partitioning and asynchronous computation by interleaving scatter-gather phases.
The 2-phased
SpMSpV algorithm 
of \cite{spmspv}
partitions the rows of matrix and buckets the non-zero entries of SpV corresponding to each partition individually.
It keeps a constant number of partitions to ensure work efficiency. However, for large graphs, 
the number of vertices in 
a partition can outgrow the cache capacity, resulting in high cache miss ratio. Unlike 
\cite{spmspv}, GPOP caps the partition size to ensure cacheability of vertices and still achieves work-efficiency using the 2-level Active List. Moreover, as explained in section \ref{sec:isg}, it is not possible to mix bucketing with the aggregation phase in \cite{spmspv}.

Another key difference between other shared-memory frameworks using GAS (with or without partitioning) 
 \cite{xstream, graphmat, hpec, spmv, spmspv} and GPOP is communication volume and memory efficiency. The former typically generate messages for neighboring {\it vertices} and propagate them in a Source-centric fashion. 
In contrast, GPOP {\it aggregates} messages to  neighboring {\it partitions} and optimally switches between dual communication modes, thus reducing DRAM traffic while simultaneously increasing bandwidth.

The vertex ID memoization used in PC mode of GPOP has previously been employed by some distributed-memory  frameworks as well \cite{gluon, slotacase}.
However, we note that shared-memory communication optimization objectives and techniques are quite distinct from distributed-memory. 
Performance in GPOP and \cite{pcpm} comes primarily from avoiding cache misses and achieving high DRAM bandwidth by enforcing a specific order of message writing (in PC mode), even if it costs additional explicit loads and stores. On the contrary, distributed-memory optimizations primarily target reduction of communication volume. All messages are locally aggregated and transmitted in bulk, and hence, message generation order is of little importance in distributed-memory frameworks. 

Some frameworks trade off programmability and preprocessing overheads for performance. Polymer \cite{polymer} is a NUMA-aware framework that reduces random accesses to remote nodes, generated by its underlying Ligra engine.
However, in certain scenarios, NUMA-aware partitioning has been shown to worsen the performance for some algorithms such as BFS \cite{everything}. Grazelle \cite{grazelle} deploys NUMA optimizations along with vectorized loop processing to improve performance of \textit{pull} direction VC. While we do not focus on these aspects 
in this paper, 
NUMA awareness can be instilled in GPOP because of the explicit communication and high access locality, and sequential accesses to bins can be vectorized. The programming interface of Polymer and Grazelle is also quite complex and requires hundreds of lines of code to be written for simple applications, such as BFS. 

Domain Specific Languages~(DSL) are another line of research in high performance graph analytics. Green-Marl \cite{greenmarl} is a popular DSL that allows users to write graph algorithms intuitively and relies on its compiler to extract parallelism and performance. GraphIt \cite{graphit} allows user to explore the trade off space of various optimizations (push vs pull, dense vs sparse frontier, NUMA partitioning etc). Perhaps, the optimizations imbibed in GPOP can also become a part of such languages in the future.

	
 \section{Conclusion}\label{sec:conclusion}
In this paper, we presented the GPOP framework for graph processing, which achieves high cache and memory performance while guaranteeing theoretical efficiency. It further utilizes the shared-memory model of target multicore platforms to enable asynchronous update propagation and faster convergence. 


We compared GPOP against Ligra, GraphMat and Galois for various graph algorithms on large datasets. GPOP incurred dramatically less L2 cache misses than the baselines, outperforming them in most of the cases. We found that DRAM bandwidth for sequential accesses is a limiting factor for GPOP's performance. Hence, execution can be further accelerated by using high bandwidth memory platforms, such as Hybrid Memory Cube. The deterministic caching capabilities of GPOP can also benefit future architectures with controllable caches or scratchpads.

\bibliographystyle{plain}
\bibliography{pcpm} 

\end{document}